\documentclass[traditabstract]{aa} 

\usepackage{amsmath}
\usepackage{graphicx}
\usepackage{txfonts}
\usepackage{natbib}
\usepackage{longtable}
\usepackage{multirow}
\newcommand{\MC}{\text{971-{\rm XMM}}}
\newcommand{\fluxunits}{{\rm erg}\;{\rm s}^{-1}{\rm cm}^{-2}}
\newcommand{\kbol}{k_{\rm bol}}
\newcommand{\Lbol}{L_{\rm bol}}
\newcommand{\Lir}{L_{\rm IR}}
\newcommand{\Lx}{L_{\rm X}}
\newcommand{\Lhard}{L_{[2-10]{\rm keV}}}
\newcommand{\Log}{{\rm Log}~}
\newcommand{\NH}{N_{\rm H}}
\newcommand{\lambdaEdd}{\lambda_{\rm Edd}}

\newcommand{\rev}[1]{{ #1}}

\begin{document}
%
   \title{The bolometric output and host-galaxy properties of obscured AGN in the XMM-COSMOS survey}

   
   \author{E. Lusso$^{1,2}$\thanks{elisabeta.lusso2@unibo.it},
          A. Comastri$^{2}$, C. Vignali$^{1,2}$, G. Zamorani$^{2}$, E. Treister$^{3,9}$, D. Sanders$^{3}$, M. Bolzonella$^{2}$, A. Bongiorno$^{4}$, M. Brusa$^{4}$, F. Civano$^{6}$, R. Gilli$^{2}$, V. Mainieri$^{7}$, P. Nair$^{2}$, M. C. Aller$^{8}$, M. Carollo$^{8}$, A. M. Koekemoer$^{11}$, A. Merloni$^{4,5}$, and J. R. Trump$^{10}$.
       }

   \institute{$^{1}$Dipartimento di Astronomia, Universit\`{a} di Bologna, via Ranzani 1, I-40127 Bologna, Italy.\\
$^{2}$INAF--Osservatorio Astronomico di Bologna, via Ranzani 1, I-40127 Bologna, Italy.\\
$^{3}$Institute for Astronomy, 2680 Woodlawn Drive, University of Hawaii, Honolulu, HI 96822, USA.\\
$^{4}$Max Planck Institut f\"{u}r extraterrestische Physik, Giessenbachstrasse 1, 85748 Garching, Germany.\\
$^{5}$Excellence Cluster Universe, TUM, Boltzmannstr.  2, D-85748, Garching bei M\"{u}nchen, Germany.\\
$^{6}$Harvard-Smithsonian Center for Astrophysics, 60 Garden Street,Cambridge, MA 02138, USA.\\
$^{7}$ESO, Karl-Schwarzschild-Strasse 2, 85748 Garching bei M\"{u}nchen, Germany.\\
$^{8}$ETH Z\"{u}rich, Physics Department, CH-8093, Z\"{u}rich, Switzerland.\\
$^{9}$Universidad de Concepci\'{o}n, Departamento de Astronom\`{i}a, Casilla 160-C, Concepci\'{o}n, Chile.\\
$^{10}$University of California Observatories/Lick Observatory, University of California, Santa Cruz, CA 95064.\\
$^{11}$Space Telescope Science Institute, Baltimore, Maryland 21218, USA.\\
}

\authorrunning{E. Lusso et al}

   \date{Accepted version August 24, 2011} 

  \abstract{We present a study of the multi-wavelength properties, from the mid-infrared to the hard X--rays, of a sample of 255 spectroscopically identified X--ray selected Type-2 AGN from the XMM-COSMOS survey. Most of them are obscured the X--ray absorbing column density is determined by either X--ray spectral analyses (for the 45\% of the sample), or from hardness ratios. Spectral Energy Distributions (SEDs) are computed for all sources in the sample. The average SEDs in the optical band is dominated by the host-galaxy light, especially at low X--ray luminosities and redshifts. There is also a trend between X--ray and mid-infrared luminosity: the AGN contribution in the infrared is higher at higher X--ray luminosities. We calculate bolometric luminosities, bolometric corrections, stellar masses and star formation rates (SFRs) for these sources using a multi-component modeling to properly disentangle the emission associated to stellar light from that due to black hole accretion. For 90\% of the sample we also have the morphological classifications obtained with an upgraded version of the Zurich Estimator of Structural Types (ZEST+). We find that on average Type-2 AGN have lower bolometric corrections than Type-1 AGN. Moreover, we confirm that the morphologies of AGN host-galaxies indicate that there is a preference for these Type-2 AGN to be hosted in bulge-dominated galaxies with stellar masses greater than $10^{10}$ solar masses. 
}
   \maketitle
%

\section{Introduction}
The formation and growth of supermassive black holes (SMBHs) and their host-galaxies are related processes. This is supported by various observational signatures: the SMBH mass correlates with the mass of the bulge of the host-galaxy (\citealt{1998AJ....115.2285M,2003ApJ...589L..21M}), with the velocity dispersion of the bulge (\citealt{2000ApJ...539L...9F,2002ApJ...574..740T}), and with the luminosity of the bulge (\citealt{1995ARA&A..33..581K}).
From theoretical models, AGN seem to be able to switch off cooling flows in clusters (\citealt{2004ApJ...617..896H}) and star formation in galaxies (\citealt{2008MNRAS.391..481S}), with the result that the SMBH mass is related to the host-galaxy bulge mass (or vice-versa). Feedback between an accreting SMBH and the host-galaxy may play an important role in galaxy formation and evolution. Understand the role of feedback is a demanding problem for both observers and theorists. Semi-analytical models and hydrodynamical simulations have been developed to attempt to link the formation and evolution of SMBHs to the structure formation over cosmic time. These models invoke different mechanisms to fuel the central SMBHs and to build the host-galaxy bulges, such as major/minor mergers of galaxies (e.g., \citealt{2000ApJ...536L..73C,2000MNRAS.311..576K,2005MNRAS.361..776S,2006ApJS..163....1H}), smooth accretion of cold gas from filamentary structures (e.g., \citealt{2009MNRAS.395..160K,2009ApJ...703..785D}), or accretion of recycled gas from dying stars (e.g., \citealt{2010ApJ...717..708C}).
Several works also consider radiative feedback which can reproduce two important phases of galaxy evolution, namely an obscured-cold-phase, when the bulk of star formation and black hole accretion occurs, and the following quiescent hot phase in which accretion remains highly sub-Eddington and unobscured (e.g., \citealt{2005MNRAS.358..168S,2011A&A...525A.115L}). 
In some of these models, the obscured/unobscured AGN dichotomy is more related to two different phases of galaxy evolution (\citealt{2008ApJS..175..356H}), rather than to an orientation effect (i.e., unified model scheme).
\par
The obscured/unobscured time dependent AGN dichotomy could be related to the bimodality in the rest-frame color distribution of host-galaxies (\citealt{2007A&A...475..115R,2007ApJ...660L..11N,2009A&A...507.1277B,2009ApJ...696..396S}), namely the red-sequence (or ``red-cloud'') and blue-cloud galaxies.
Broad-line AGN (if the morphology of the host-galaxy is available) are likely to be associated to galaxies belonging to the blue-cloud, while obscured objects to red passive galaxies. The green valley should be populated by transition objects.
The picture above is probably a too crude approximation. 
Moreover, one should note that red sequence galaxies may well be passively evolving galaxies without significant star formation (e.g., \citealt{2007A&A...475..115R,2007ApJ...660L..11N}), rather than dusty starforming objects (e.g., \citealt{2009A&A...507.1277B}).
\par
Disentangling the contribution of the nuclear AGN from the host-galaxy properties in the broad band SED is fundamental to constrain the physical evolution of AGN and to place them into the context of galaxy evolution.
In the standard picture the AGN energy output is powered by accretion onto SMBHs. The disk accretion emission is visible in the optical-UV as the blue-bump feature. The X--ray emission is believed to be due to a hot-electrons corona that surrounds the accretion disk, while the infrared emission is likely due to the presence of a dusty torus around the disk at few parsec from the center, which reprocesses the nuclear radiation.
According to the unified model of AGN (e.g., \citealt{1993ARA&A..31..473A,1995PASP..107..803U}), hot dust is located in the inner edge of the torus.
However, recent studies predict and observe exceptions to the unified model. 
From the theoretical point of view, an alternative solution to the torus is the disk-wind scenario (e.g., \citealt{1992ApJ...385..460E,2006ApJ...648L.101E}). From the observational side, AGN without any detectable hot dust emission (e.g., \citealt{2010Natur.464..380J}) and weak infrared emission (e.g., \citealt{2010ApJ...724L..59H}) are predicted and observed.
The vast majority of studies performed so far concern unobscured (Type-1) AGN for which their SED is well known from low-$z$ ($\langle z \rangle\sim0.206$, see \citealt{1994ApJS...95....1E}) to high-$z$ ($\langle z \rangle\sim1.525$, see \citealt{2006ApJS..166..470R}).
An obvious complication in the study of their host-galaxy properties is that the emission of the central AGN outshines the galaxy light in UV, optical and infrared bands; therefore it is extremely difficult to derive constraints on the colors, stellar populations, and morphologies of the host.
On the other hand, for obscured (Type-2) AGN the host-galaxy light is the dominant component in the optical/near-infrared SED, while it is difficult to recover the AGN intrinsic nuclear emission. The lack of a proper characterization of the nuclear componentof the SED of obscured Type-2 AGN is a major limitation. 
As a consequence, the relations between stellar masses, SFR, morphologies and accretion luminosity remain poorly known.
Since the relative contribution in the SED of the different components (AGN/host-galaxy) varies with wavelength, a proper decomposition can be obtained by an SED-fitting approach, complemented by a morphological analysis. 
This will provide a robust estimate of the nuclear emission (bolometric luminosities and bolometric corrections, absorption column density distributions, etc) and its relation with the host-galaxy properties (mass, star formation rates, morphological classification). 
The AGN structure is reflected in the shape of the SED, specifically the big-blue bump and the infrared-bump are related to the accretion disk and the surrounding torus, respectively. Therefore, a densely sampled SED over a broad wavelength interval is mandatory to extract useful information from SED fitting procedures, allowing to tightly constrain physical parameters from multi-component modeling and, in particular, to properly disentangle the emission associated to stellar light from that due to accretion.
\par
The combination of sensitive X--ray and mid-IR observatories allows us to model the obscuring gas that in Type-2 AGN hides the nuclear region from the near-IR to the UV. 
As supported by previous investigations, the reprocessed IR emission could be a good proxy of the intrinsic disk emission.
\citet{2009A&A...502..457G} confirm the correlation between the X--ray luminosity at [2-10]keV and the IR emission for a sample of Seyfert galaxies (see also \citealt{2004A&A...418..465L}).
Their data are the best estimate of the nuclear (non stellar) IR flux in AGN to date. A highly significant correlation between $\Lhard$ and the intrinsic nuclear IR luminosity at 12.3$\mu$m is observed in the high quality near-IR and X--ray data discussed by \citet{2009A&A...502..457G}.
This reinforces the idea that ``uncontaminated'' mid-IR continuum is an accurate proxy for the intrinsic AGN emission.
\par
In this work we present the largest study of the multi-wavelength properties of an X--ray selected sample of obscured AGN using the XMM-Newton wide field survey in the COSMOS field (XMM-COSMOS). Following a similar approach to that of \citet{2007A&A...468..603P} and \citet{2010MNRAS.402.1081V}, we use the infrared emission to evaluate the nuclear bolometric luminosity from a multi-component fit.
The paper is aimed at a detailed characterization of a large sample of obscured AGN over a wide range of frequencies. The SEDs, morphology of the host-galaxies, stellar masses, colors, bolometric luminosities and bolometric corrections for the sample of obscured AGN are presented.
\par
This paper is organized as follows. In Sect.~\ref{The Data Set} we report the selection criteria for the sample used in this work. Section~\ref{Rest-frame monochromatic fluxes and Spectral Energy Distributions} presents the multi-wavelength data-set, while in Sect.~\ref{Average SED} the method to compute average SED is described. Section~\ref{SED-fitting} concerns the multi-component modeling used to disentangle the nuclear emission from the stellar light. In Section~\ref{Bolometric luminosities and bolometric corrections} the method used to compute intrinsic bolometric luminosites and bolometric corrections for Type-2 AGN is described, while in Sect.~\ref{Calibrating the method} we have applied the same method to a sample of Type-1 AGN.
The discussion of our findings is given in Sect.~\ref{Results and discussion}, while in Sect.~\ref{Summary and Conclusions} we summarize the most important results.
\par
We adopted a flat model of the universe with a Hubble constant $H_{0}=70\, \rm{km \,s^{-1}\, Mpc^{-1}}$, $\Omega_{M}=0.27$, $\Omega_{\Lambda}=1-\Omega_{M}$ (\citealt{komatsu09}).

\section{The Data Set}
\label{The Data Set}
The XMM-COSMOS catalog comprises $1822$ point--like X--ray sources detected by XMM-\textit{Newton} over an area of $\sim 2\,\rm deg^2$. The total exposure time was $\sim 1.5$ Ms with a fairly homogeneous depth of $\sim 50$ ks over a large fraction of the area (\citealt{hasinger07}, \citealt{cappelluti09}).
Following \citet{2010ApJ...716..348B}, we excluded from our analysis 25 sources which turned out to be a blend of two {\it Chandra} sources. This leads to a total of 1797 X--ray selected point-like sources. 
We restricted the analysis to 1078 X--ray sources detected in the [2-10]~keV band at a flux larger than $2.5\times10^{-15}\fluxunits$ (see Table 2 in \citealt{cappelluti09}). The objects for which no secure optical counterpart could be assigned are often affected by severe blending problems, so that we consider in this analysis the 971 sources (hereafter $\MC$) for which a secure optical counterpart can be associated (see discussion in \citealt{2010ApJ...716..348B})\footnote{The multi-wavelength XMM-COSMOS catalog can be retrieved from: http://www.mpe.mpg.de/XMMCosmos/xmm53\_release/, version $1^{\rm st}$ April 2010.}.
\par
From the $\MC$ catalog we have selected $255$ sources, which do not show broad (FWHM$<2000$ km s$^{-1}$) emission lines in their optical spectra\footnote{The origin of spectroscopic redshifts for the $255$ sources is as follows: $11$ objects from the SDSS archive, $2$ from MMT observations (\citealt{prescott06}), $70$ from the IMACS observation campaign (\citealt{trump07}), $156$ from the zCOSMOS bright $20$k sample (see \citealt{lilly07}), $7$ from the zCOSMOS faint catalog and 9 from the Keck/DEIMOS campaign.} (hereafter we will refer to them as the Type-2 AGN sample): 223 are classified not-broad-line AGN, while 32 are absorption-line galaxies.
Not-broad-line AGN are objects with unresolved, high-ionization emission lines, exhibiting line ratios indicating AGN activity, and, when high-ionization lines are not detected, or the observed spectral range does not allow to construct line diagnostics, objects without broad line in the optical spectra. Absorption-line galaxies are sources consistent with a typical galaxy spectrum showing only absorption line. 
 
\begin{figure}
 \includegraphics[width=8cm]{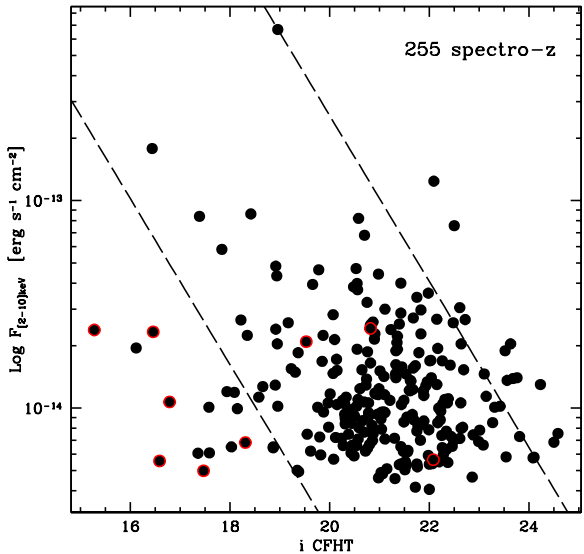}
 \caption{Plot of the $[2-10]$keV flux versus the total $i^*$ CFHT magnitude for our sample of 255 Type-2 AGN. The red circles represent sources with a de-absorbed 2--10 keV luminosity lower than $10^{42}$ erg s$^{-1}$. The dashed lines represent a constant X--ray to optical flux ratio of $\Log(X/O)=\pm 1$.}
 \label{fluxMHi}
\end{figure}
In Figure \ref{fluxMHi} we plot the 2--10 keV X--ray flux as a function of $i^*$ CFHT magnitude. The dashed lines limit the region typically occupied by AGN along the X--ray to optical flux ratio $\Log(X/O)=\pm1$\footnote{$\Log(X/O)=\Log f_x+i^*/2.5+5.6$.}. Nine sources have a de-absorbed 2--10 keV luminosity lower than $10^{42}$ erg s$^{-1}$, the conventional threshold below which the X--ray sources that can plausibly be explained by moderate-strength starbursts, hot gas in elliptical galaxies, or other sources besides accretion onto a nuclear SMBH (\citealt{2001ApJ...554..742H}).
The three sources inside the dashed lines have X--ray luminosities close to $10^{42}$ erg s$^{-1}$, while six AGN (6/255, 2\%) lie in the part of the diagram usually occupied by star-forming galaxies, and have X--ray luminosities $<10^{42}$ erg s$^{-1}$. Their inclusion in the analysis does not affect the main results.

\begin{figure*}
  \centering
  {\label{histredshift}\includegraphics[width=0.3\textwidth]{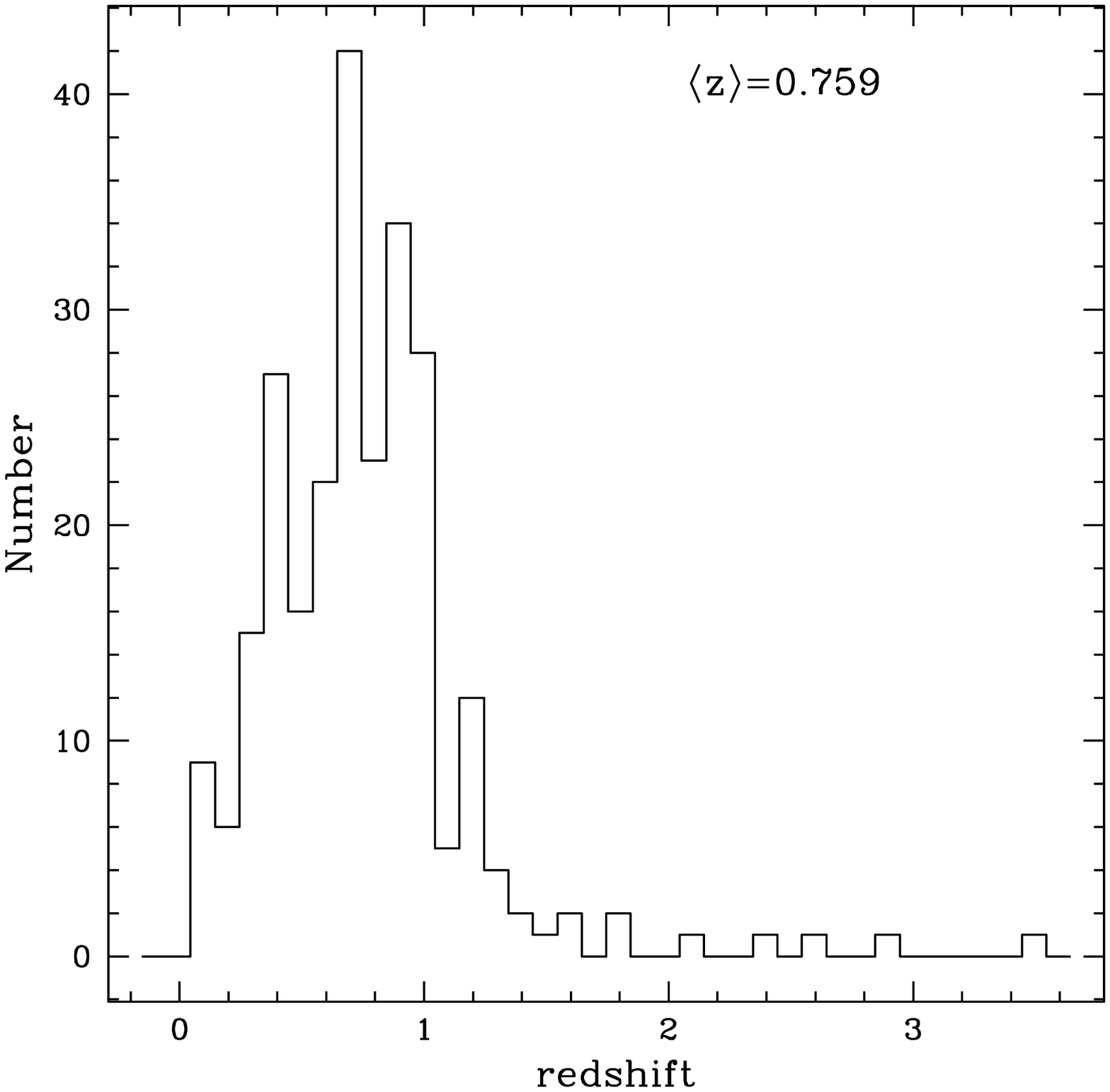}}                
  {\label{histlx}\includegraphics[width=0.3\textwidth]{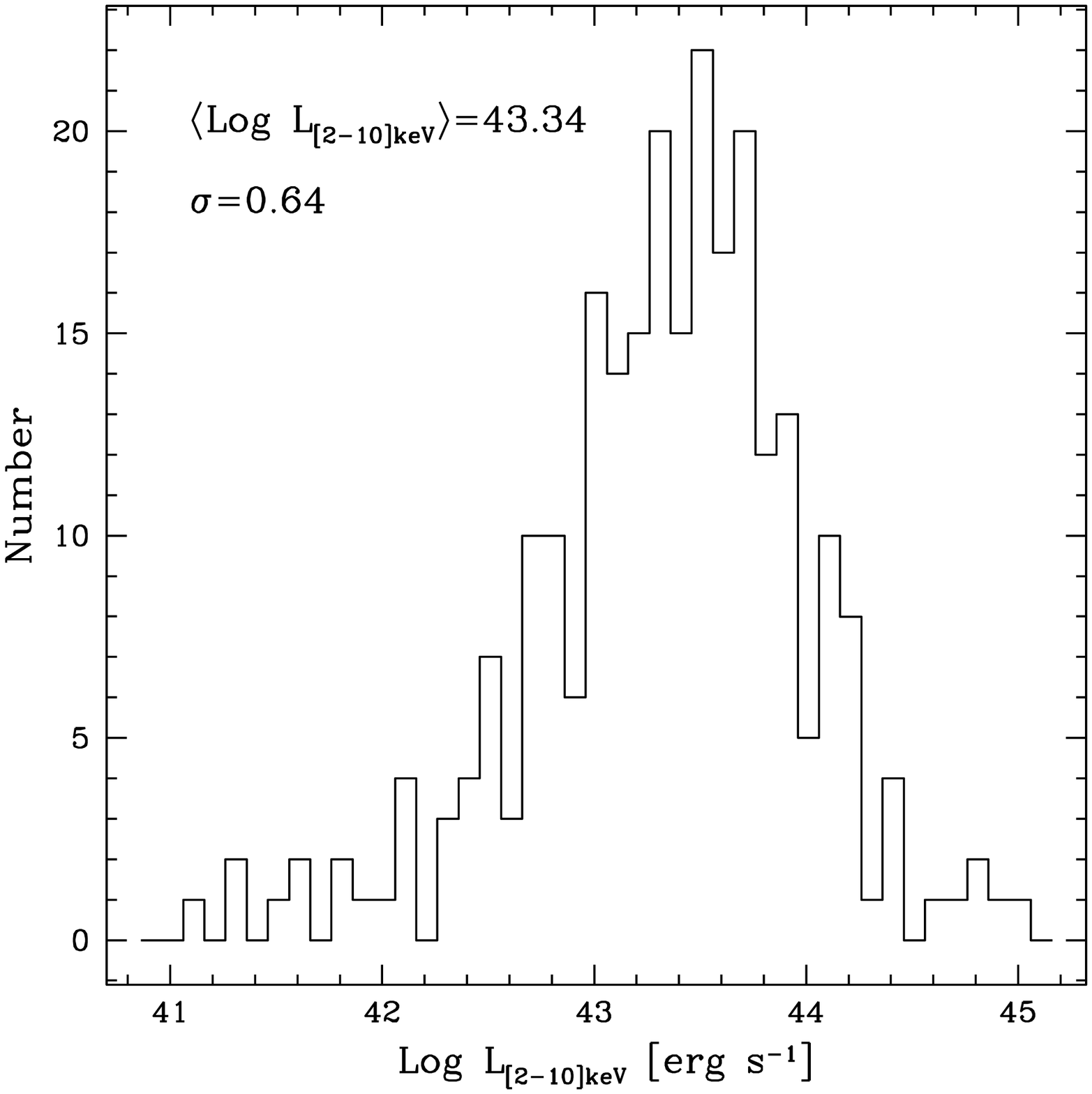}}
  {\label{histnh}\includegraphics[width=0.3\textwidth]{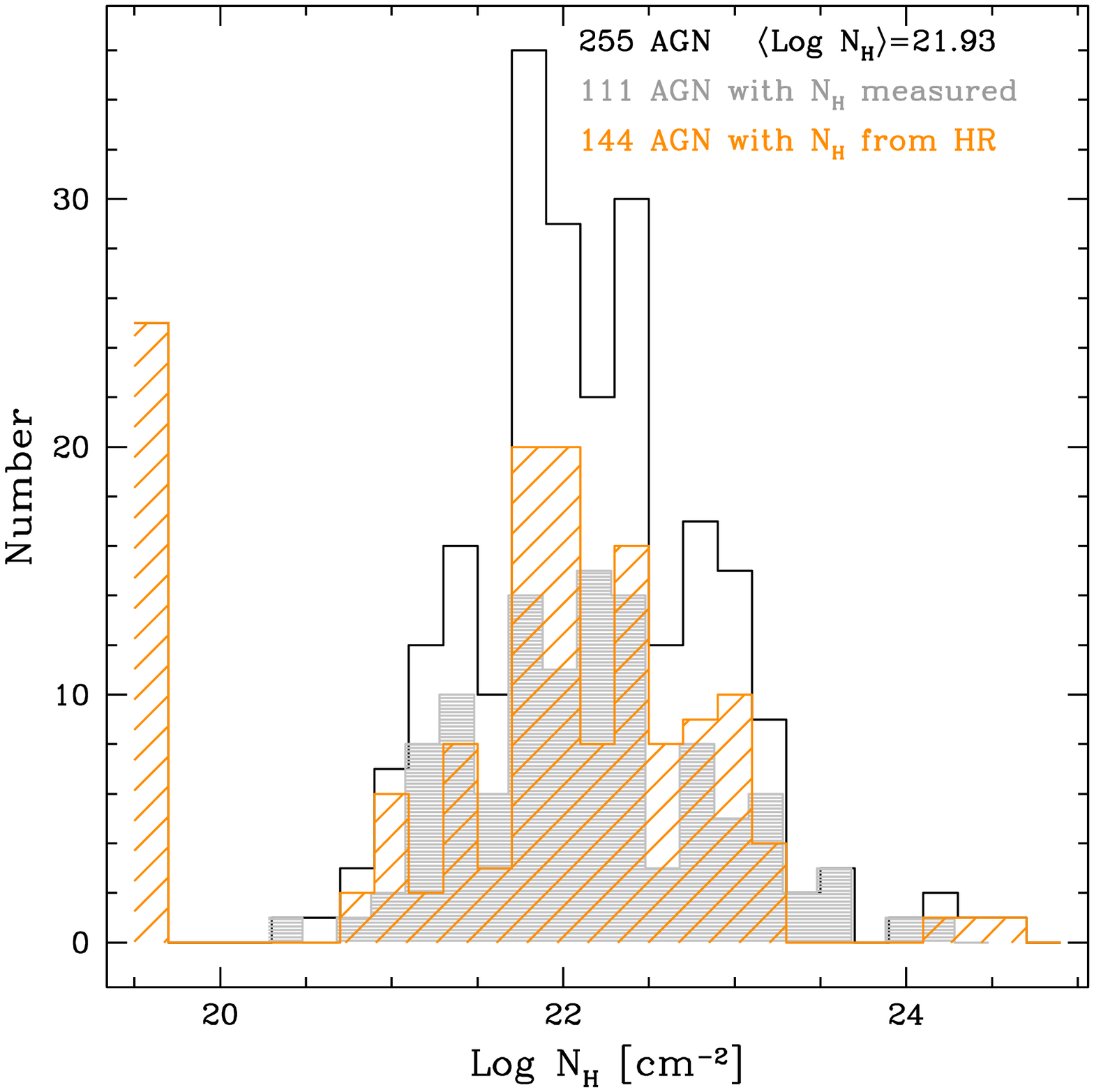}}
  \caption{\textit{Left panel:} Redshift distribution of the $255$ Type-2 AGN considered in this work. \textit{Center panel:} Intrinsic hard X--ray luminosity distribution of the $255$ Type-2 AGN considered in this work. \textit{Right Panel:} Column density distribution of the 255 Type-2 AGN (\textit{black histogram}), of the 111 Type-2 AGN with an $\NH$ estimate from spectral analysis (\textit{grey filled histogram}), and of the 144 Type-2 AGN with an $\NH$ estimate from hardness ratios (\textit{orange hatched histogram}).}
  \label{fig:zlxnh}
\end{figure*}

The Type-2 AGN sample used in our analysis comprises 255 X--ray selected AGN, all of them with spectroscopic redshifts, spanning a wide range of redshifts ($0.045<z<3.524$) and X--ray luminosities ($41.06 \leq \Log \Lhard \leq 45.0$).
The redshift distribution of the total sample and the distribution of the de-absorbed hard X--ray luminosities are presented in Figure \ref{histredshift}. The mean redshift is $\langle z\rangle=0.76$, while the mean $\Log \Lhard$ is 43.34 with a dispersion of 0.64.

\subsection{Correction for absorption for the X--ray luminosity}
\label{Correction for absorption for the X--ray 2--10 keV luminosity}
For a sub-sample of 111 AGN we have an estimate of the column density $N_{\rm H}$ from spectral analysis (see \citealt{2007ApJS..172..368M}), while for 144 AGN absorption is estimated from hardness ratios (HR; see \citealt{2010ApJ...716..348B}). For 25 sources for which a column density estimate is not available from HR we consider the Galactic column density. Therefore, we can compute the de-absorbed X--ray luminosity at 0.5--2 keV (soft band) and 2--10 keV (hard band) for all sources in our sample. In Figure \ref{histnh} (right panel) we show the distribution of column densities that ranges from $3\times10^{20}$ cm$^{-2}$ to $1.5\times10^{24}$ cm$^{-2}$. The mean $N_{\rm H}$ value is $8.5\times10^{21}$ cm$^{-2}$ with a dispersion of 0.72 dex.
The integrated intrinsic un-absorbed luminosity is computed assuming a power-law spectrum with slope, $\Gamma=2$ and $\Gamma=1.7$ for the 0.5--2 keV and 2--10 keV bands, respectively.
The average shift induced by the correction for absorption in the Type-2 sample is $\langle\Delta \Log\Lhard\rangle=0.04\pm0.01$.


\section{Rest-frame monochromatic fluxes and \\Spectral Energy Distributions}
\label{Rest-frame monochromatic fluxes and Spectral Energy Distributions}
We used the catalog by \citet{2010ApJ...716..348B} which includes multi-wavelength data from mid-infrared to hard X--rays: MIPS 160 $\mu$m, 70 $\mu$m and 24 $\mu$m GO3 data (\citealt{2009ApJ...703..222L}), IRAC flux densities (\citealt{sanders07}), $u^*$ and $i^*$ CFHT bands and near-infrared $K_S$-band data (\citealt{mccraken08}), J UKIRT (\citealt{2008yCat.2284....0C}), HST/ACS F814W imaging of the COSMOS field (\citealt{koekemoer07}), optical multiband photometry (SDSS, Subaru, \citealt{capak07}) and near- and far-ultraviolet bands with GALEX (\citealt{2007ApJS..172..468Z}).

The number of X--ray sources detected at $160\,\mu m$ and $70\,\mu m$ is 18 and 42, respectively. For the undetected sources in these bands we consider 5$\sigma$ upper limits of $65 \, \rm m Jy$ and $8.5 \, \rm m Jy$ for $160\,\mu m$ and $70\,\mu m$, respectively. At $24\,\mu m$ the number of detected sources is 237. For the 18 undetected sources at $24\,\mu m$, we consider 5$\sigma$ upper limits of $80 \, \rm \mu Jy$. All 255 sources are detected in the infrared in all IRAC bands, and only a few objects were not detected in the optical and near-IR bands: we have only 8 upper limits in the $z^+$ band; 1 upper limit in the $B_{J}$ and $i^*$ bands; 2 upper limits in the $u^{*}$ band; 4 upper limits in the $K_{S}$ CFHT band and 2 in the $J$ UKIRT band. The observations in the various bands are not simultaneous, as they span a time interval of about 5 years: 2001 (SDSS), 2004 (Subaru and CFHT) and 2006 (IRAC). 
Variability for absorbed sources is likely to be a negligible effect, but, in order to further reduce it, we selected the bands closest in time to the IRAC observations (i.e., we excluded SDSS data, that in any case are less deep than other data available in similar bands). GALEX bands are not taken into account because, given the large aperture they can include light from close companions.
All the data for the SED computation were shifted to the rest frame, so that no K-corrections were needed.
Galactic reddening has been taken into account: we used the selective attenuation of the stellar continuum $k(\lambda)$ taken from Table 11 of \citet{capak07}. Galactic extinction is estimated from \citet{schlegel98} for each object in the $\MC$ catalog. Count rates in the 0.5-2 keV and 2-10 keV are converted into monochromatic X--ray fluxes in the observed frame at 1 and 4 keV, respectively, using a Galactic column density $\NH = 2.5 \times 10^{20}\,cm^{-2}$ (see \citealt{cappelluti09}), and assuming a photon index $\Gamma_x=2$ and $\Gamma_x=1.7$, for the soft and hard band, respectively. We do not correct these X--ray fluxes for the intrinsic column density. All sources are detected in the 2-10 keV band by definition of the sample, while in the soft band we have 70 upper limits. 

\section{Average SED}
\label{Average SED}
\begin{figure}
 \includegraphics[width=8cm]{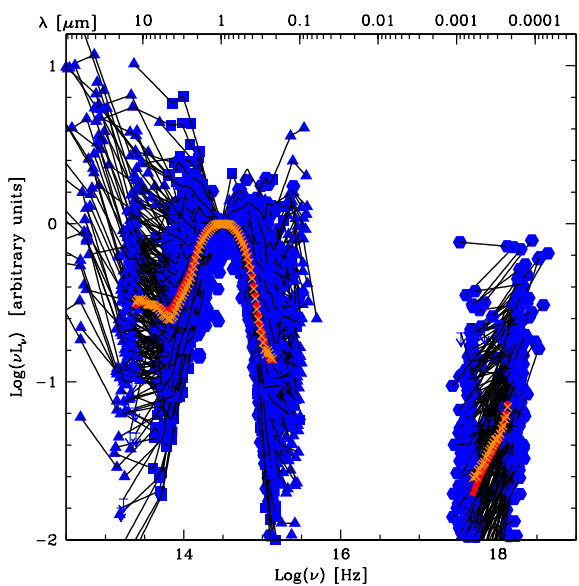}
 \caption{Mean (\textit{orange crosses}) and median (\textit{red points}) SED from the total sample of 255 Type-2 AGN. The blue points represent the rest-frame data, from infrared to X--ray, used to construct the average SED, while the black lines represent the interpolated SED.}
 \label{meansed257}
\end{figure}
We have computed the individual rest-frame SEDs for all sources in the sample, following the same approach as in L10, and we have normalized all of them at 1$\mu$m. 
After this normalization we divided the frequency interval from $\Log \nu_{\rm min}$ to $\Log \nu_{\rm max}$ using a fixed step $\Delta \Log \nu$. The minimum and maximum frequency depends on both the data and the redshift of the source considered to compute the SED, in our case we have used $\Log \nu_{\rm min}=12$ Hz and $\Log \nu_{\rm max}=20$ Hz, with a $\Delta \Log \nu=0.02$.
We averaged data in each given interval $\Log \nu_{l} \leq \Log \nu \leq \Log \nu_{ l+1}$. 
The mean and median SEDs are obtained by taking the arithmetic mean and the median of logarithmic luminosities, $\Log L$, in each bin.
It is important to note that sources at different redshift contribute to the same bin.
Because of the relatively wide range of redshifts, the lowest and the highest frequency bins are populated by a variable number of points. 
This effect may introduce relatively high fluctuations in the average luminosity in those bins.
In order to minimize these effects we select a minimum number of 200 SEDs in each bin (this number depends on the total number of sources in the sample). Then, we select the mean reference frequency, $\overline{\Log \nu}$, of the bin and use a \texttt{binary-search} algorithm to find all luminosities that correspond at $\overline{\Log \nu}$ (if a source does not have a frequency that correspond to $\overline{\Log \nu}$, we choose the luminosity with the closer frequency to $\overline{\Log \nu}$).
Finally, all adjacent luminosities in each bin are then connected to compute the final mean and median SED.
In Figure \ref{meansed257} the resulting mean and median SEDs are reported with orange crosses and red points, respectively. 
The data points are reported in order to show the dispersion with respect to 1$\mu$m.
The average SED is characterized by a flat X--ray slope, $\langle\Gamma=1.12\rangle$, while in the optical-UV the observed emission appears to be consistent with the host-galaxy. The flattening of the X--ray slope is likely due to the fact that we do not correct for the intrinsic absorption the fluxes at 1 and 4 keV. The average SED in the mid-infrared is probably a combination of dust emission from star-forming region and AGN emission reprocessed by the dust. 
\begin{figure*}
  \centering
  {\label{meanbinx}\includegraphics[width=0.3\textwidth]{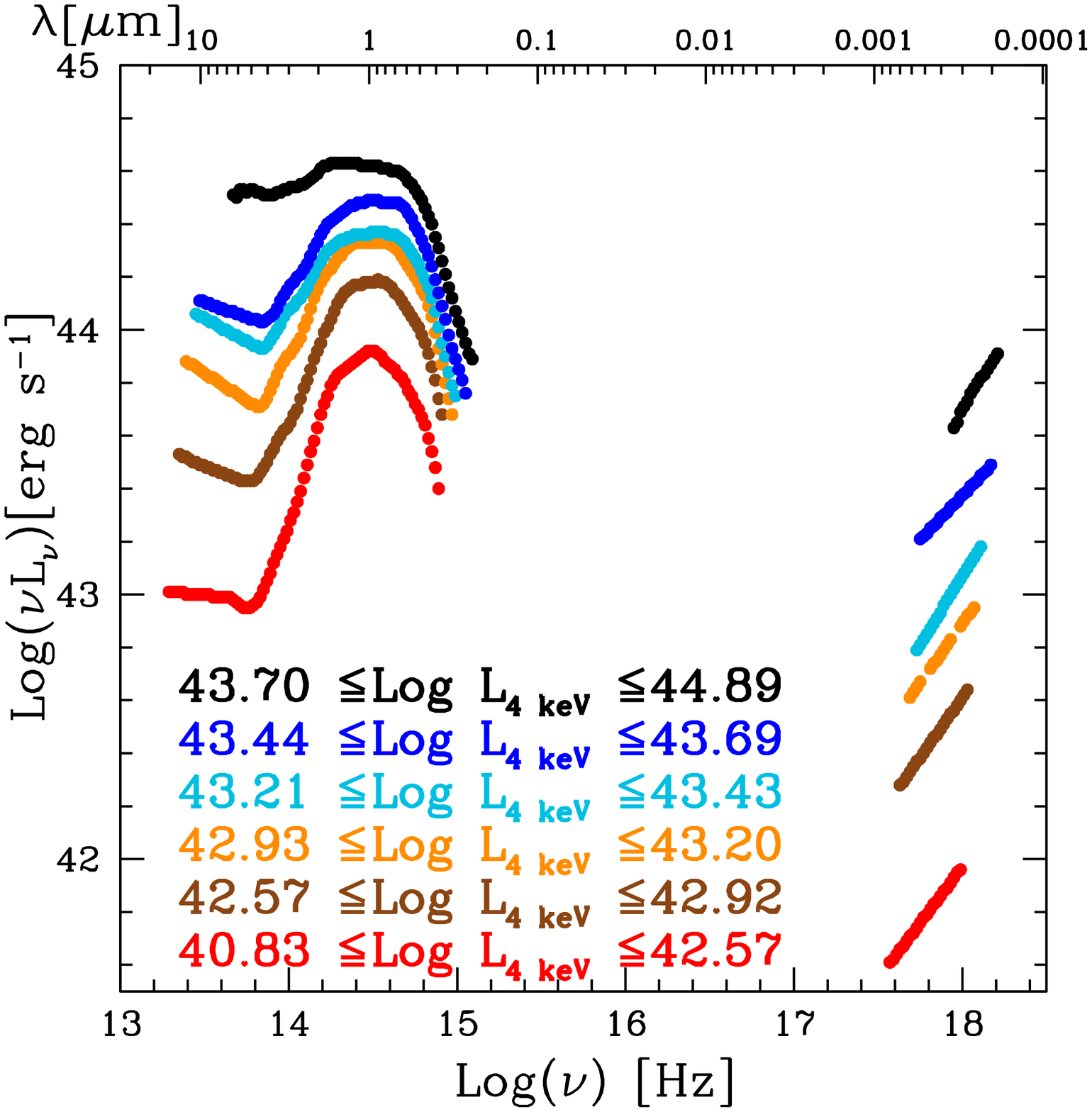}}                
  {\label{meanbinir}\includegraphics[width=0.3\textwidth]{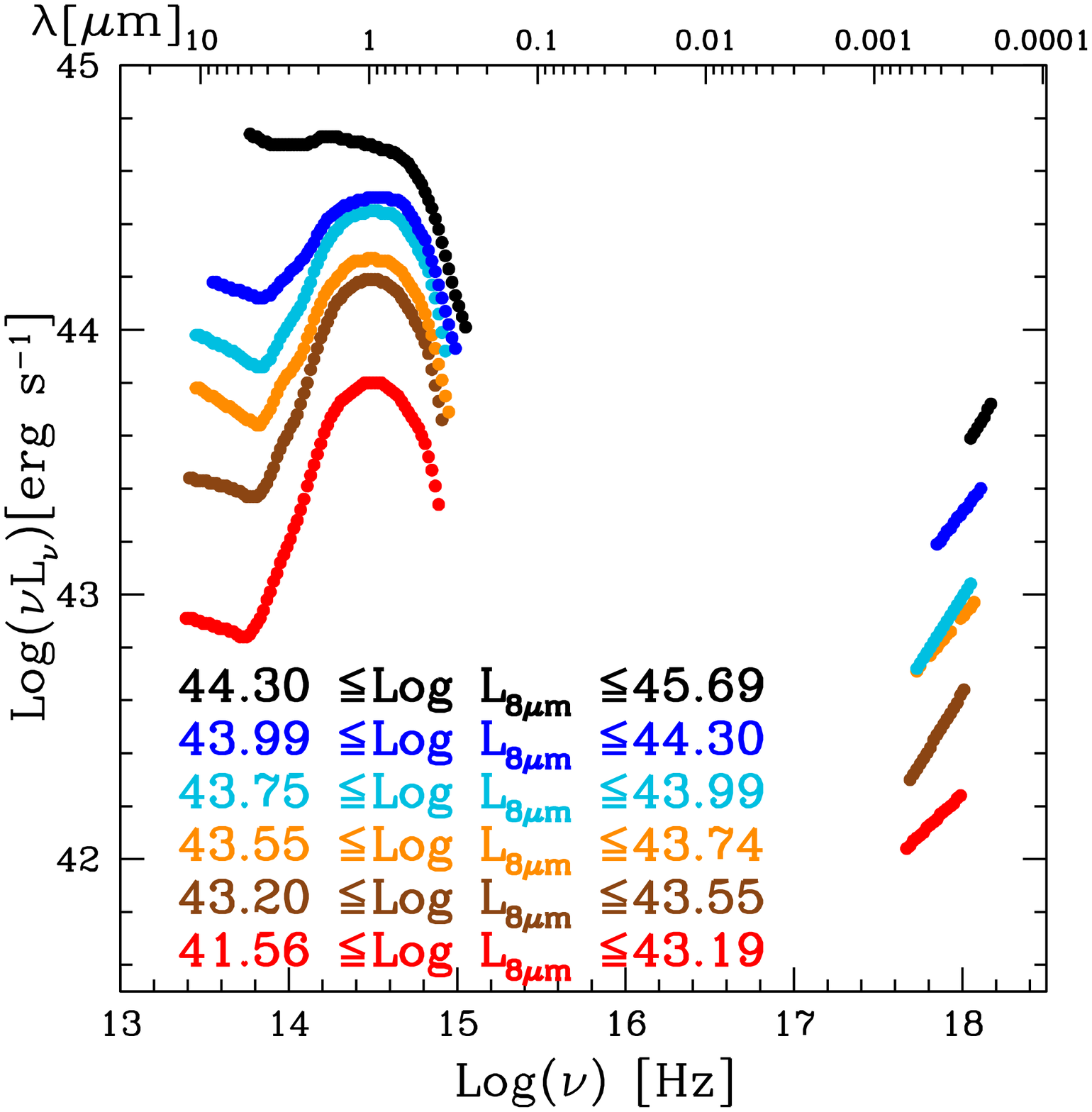}}
  {\label{meanbinz}\includegraphics[width=0.3\textwidth]{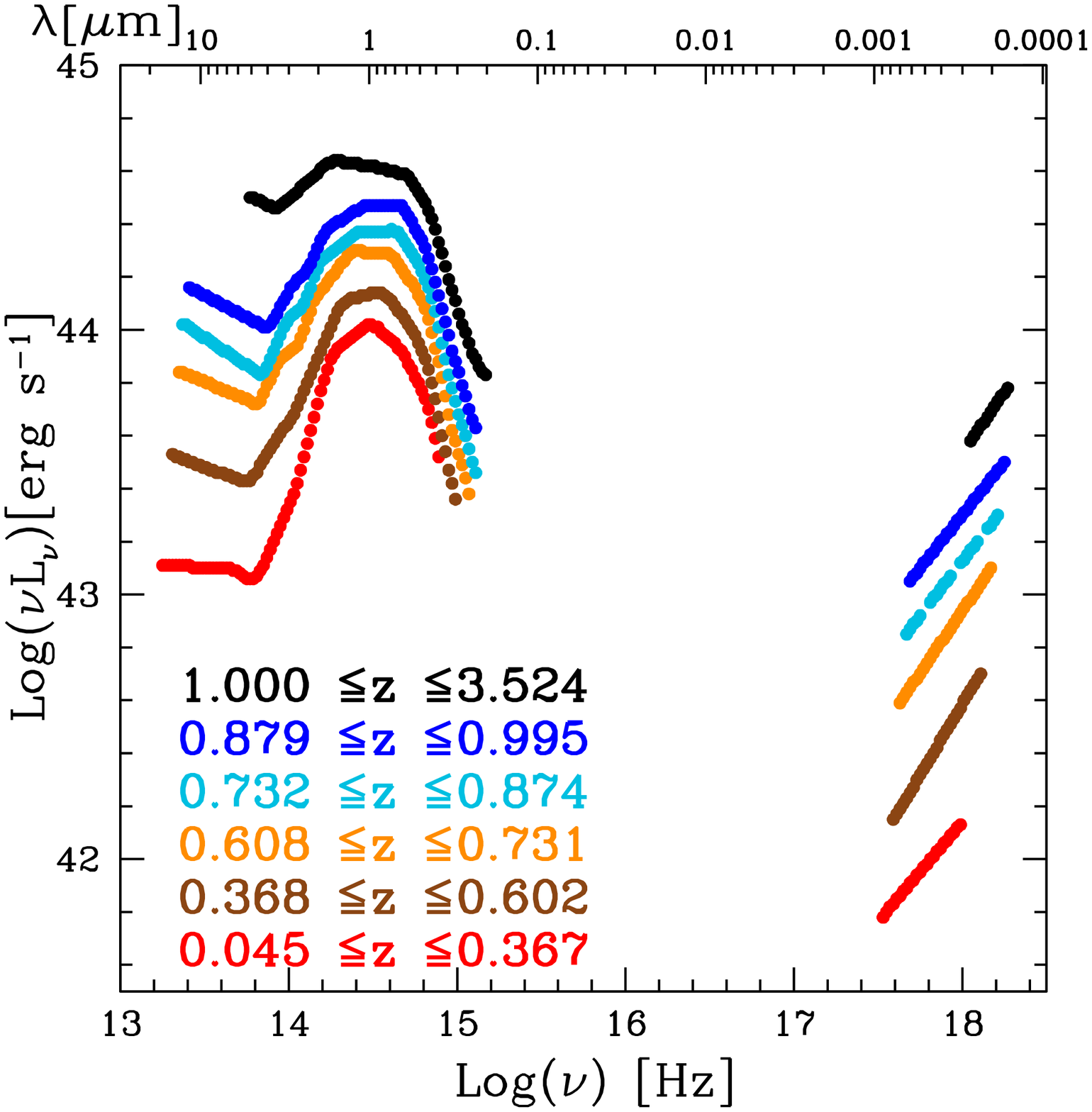}}
  \caption{Average SEDs in the rest-frame $\Log(\nu L_\nu)-\Log \nu$ plane. \textit{Right panel:} Mean SEDs computed binning in X--ray luminosity at 4 keV. \textit{Center panel:} Mean SEDs computed binning in infrared luminosity at 8$\mu$m. \textit{Left Panel:} Mean SEDs computed binning in redshift. The color code refers to the different bins as labeled.}
  \label{fig:averagesed}
\end{figure*}
Before trying to deconvolve each source using an SED-fitting code, we binned the total sample in X--ray and infrared luminosites and redshift.
We used the luminosity at 4 keV and 8$\mu$m to divide the total sample in 6 bins with the same number of sources in each bin.
The wavelength of the luminosity used to bin the sample is chosen to minimize the host-galaxy contribution.
In the three panels of Figure \ref{fig:averagesed} the resulting mean SEDs are shown.
The redshift trend is directly connected to the luminosity trend, since at higher redshifts we are looking at the more luminous sources.
The shapes of the average SEDs in the optical bands are approximately the same in all luminosity and redshift bins.
As expected, there is a stronger host-galaxy contribution at lower luminosity/redshift bins, where the average SEDs have a typical galaxy shape.
Moreover, there is a trend between X--ray and mid-infrared luminosity: the contribution in the infrared is higher at higher X--ray luminosities. 
This effect is already known for both Type-1 and Type-2 AGN using the intrinsic (non-stellar) emission from the AGN (e.g., \citealt{2004A&A...418..465L}, \citealt{2009A&A...502..457G}).
The observed average SED for our sample is a combination of both host-galaxy and AGN light, but the change in the average shape in the mid-infrared as a function of the X--ray luminosity suggests that most luminous sources are probably dominated by the AGN emission at those wavelengths.

\section{SED-fitting}
\label{SED-fitting}
\begin{figure*}
\centering
 \includegraphics[width=13cm,height=13cm]{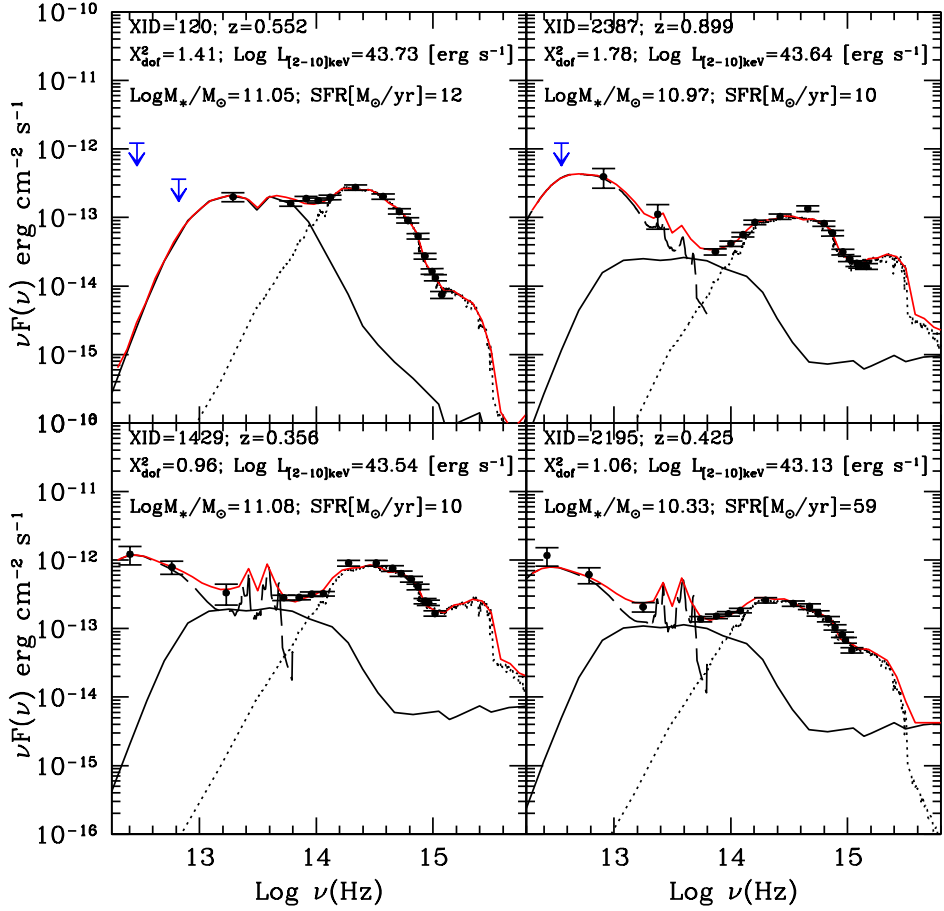}
 \caption{Examples of SED decompositions. Black circles are the observed photometry in the rest-frame (from the far-infrared to the optical-UV). The long-dashed, solid and dotted lines correspond respectively to the starburst, AGN and host-galaxy templates found as the best fit solution. The red line represents the best-fit SED. The stellar mass and the SFR derived from the galaxy template are reported.}
 \label{sed}
\end{figure*}

The purpose of performing SED fitting is to properly disentangle the emission associated to stellar light from that due to accretion and constrain physical parameters. 
Since the relative contribution of the different components varies with wavelength, a proper decomposition can be obtained through an SED-fitting approach providing a robust estimate of the nuclear emission (bolometric luminosities and bolometric corrections) and its relation with the host-galaxy properties (mass, star formation rates, morphological classification).
A well sampled SED is mandatory; in particular, far-infrared observations are fundamental to sample the star-formation activity, while mid-infrared observations are necessary to sample the region of the SED where most of the bolometric luminosity of obscured AGN is expected to be re-emitted.
In Fig.~\ref{sed} the broad-band SEDs of four XMM-Newton Type-2 AGN are plotted as examples. The lower two panels are representative of a full SED with all detections from the far-infrared to the optical. Unfortunately, there are very few detections at 160 and 70$\mu$m, so that the more representative situation is shown in the upper left panel of Fig.~\ref{sed}. The three components adopted in the SED-fitting code, \textit{starburst}, \textit{AGN torus} and \textit{host-galaxy} templates, are shown as a long-dashed line, solid line and dotted line, respectively. All the templates used in the SED-fitting code will be described in the following Sections. The red line represents the best-fit, while the black points represent the photometric data used in the code, from low to high frequency: MIPS-Spitzer ($160\mu m$, $70\mu m$ and $24\mu m$ if available), 4 IRAC bands, K CFHT, J UKIRT, optical Subaru and CFHT bands.
\par
The observed data points from infrared to optical are fitted with a combination of various SED templates (see Sect.~\ref{Template libraries}) using a standard $\chi^2$ minimization procedure
\begin{equation}
\label{chisquare}
 \chi^2=\sum_{i=1}^{\rm{n_{filters}}}\left[\frac{F_{obs,i}-A\times F_{gal,i}-B\times F_{agn,i}-C\times F_{ir,i}}{\sigma_i}\right]^2
\end{equation}
where $F_{obs,i}$ and $\sigma_i$ are the monochromatic observed flux and its error in the band $i$; $F_{gal,i}$, $F_{agn,i}$ and $F_{ir,i}$ are the monochromatic template fluxes for the host-galaxy, the AGN and the starburst component, respectively; $A$, $B$ and $C$ are the normalization constants for the host-galaxy, AGN and starbust component, respectively.
The starburst component is used only when the source is detected at $160\mu m$ and $70\mu m$. Otherwise, a two components SED-fit is used. Sixteen is the maximum number of bands adopted in the SED-fitting (only detection are considered), namely: $160\mu m$, $70\mu m$, $24\mu m$, $8.0\mu m$, $5.8\mu m$, $4.5\mu m$, $3.6\mu m$, $K_S$, $J$, $z^+$, $i^*$, $r^+$, $g^+$, $V_J$, $B_J$ and $u^*$.

\subsection{Template libraries}
\label{Template libraries}

\subsubsection{Optical template library}
We used a set of 75 galaxy templates built from the \citet[BC03 hereafter]{2003MNRAS.344.1000B} code for spectral synthesis models, using the version with the \textquotedblleft{Padova 1994}\textquotedblright~tracks, solar metallicity and Chabrier IMF (\citealt{2003ApJ...586L.133C}).
For the purposes of this analysis a set of galaxy templates representative of the entire galaxy population from ellipticals to starbursts is selected. To this aim, 10 exponentially decaying star formation histories with characteristic times ranging from $\tau = 0.1$ to $30$\,Gyr and a model with constant star formation are included.
For each SFH, a subsample of ages available in BC03 models is selected, to avoid both degeneracy among parameters and speed up the computation. In particular, early-type galaxies, characterised by a small amount of ongoing star formation, are represented by models with values of $\tau$ smaller than $1$ Gyr and ages larger than $2$\,Gyr, whereas more actively star forming galaxies are represented by models with longer values of $\tau$ and a wider range of ages from $0.1$ to $10$\,Gyr.
An additional constraint on the age is that, for each sources, the age has to be smaller than the age of the Universe at the redshift of the source.
Each template is reddened according to the \citet{2000ApJ...533..682C} reddening law. The input value is $E(B-V)$, corresponding to a dust-screen model, with $F_o(\lambda)=F_i(\lambda)10^{-0.4 E(B-V) k(\lambda)}$, where $F_o$ and $F_i$ are the observed and the intrinsic fluxes, respectively. The extinction at a wavelength $\lambda$ is related to the colour excess $E(B-V)$ and to the reddening curve $k(\lambda)$ by $A_\lambda=k(\lambda)E(B-V)=k(\lambda)A_V /R_V$, with $R_V=4.05$ for the Calzetti's law. The $E(B-V)$ values range between 0 and 1 with a step of $0.05$.

\subsubsection{AGN template library}
The nuclear SED templates are taken from \citet{2004MNRAS.355..973S}. They were constructed from a large sample of Seyfert galaxies selected from the literature for which clear signatures of non-stellar nuclear emission were detected in the near-IR and mid-IR. After a proper subtraction of the stellar contribution, the nuclear infrared data were interpolated with a radiative transfer code for dust heated by a nuclear source with a typical AGN spectrum, and including different geometries, dust distribution, variation of the radii, density and dust grain sizes to account for possible deviations from a standard ISM extinction curve (see for more details \citealt{1994MNRAS.268..235G,2001A&A...365...37M}).
\par
The infrared SEDs were then normalized by the intrinsic, unabsorbed X-ray flux in the 2--10 keV band, and are divided into 4 intervals of absorption: $\NH<10^{22}$ cm$^{-2}$ for Sy1, while $10^{22}<\NH<10^{23}$ cm$^{-2}$, $10^{23}<\NH<10^{24}$ cm$^{-2}$ and $\NH>10^{24}$ cm$^{-2}$ for Sy2 (see Fig.~1 in \citealt{2004MNRAS.355..973S}).
The main differences between the SEDs of Sy1s and Sy2s with $10^{22}<\NH<10^{23}$ cm$^{-2}$ are the absorption in the near-IR at about $\lambda<2\mu$m and the silicate absorption at $\lambda=9.7\mu$m, which are present in the Sy2 template. The shape of the SED in the mid-infrared with $10^{23}<\NH<10^{24}$ cm$^{-2}$ is quite similar to that with $10^{22}<\NH<10^{23}$ cm$^{-2}$. The Compton-thick SED ($\NH>10^{24}$ cm$^{-2}$) shows conspicuous absorption also at $\lambda\sim1.3\mu$m.
If a source has the $N_{\rm H}$ value available, this is used as a prior in the selection of the best-fit AGN template.

\subsubsection{Starburst template library}
We used two different starburst template libraries for the SED-fitting: \citet{2001ApJ...556..562C} and \citet{2002ApJ...576..159D}. These template libraries represent a wide range of SED shapes and luminosities and are widely used in the literature. Here, we briefly describe how each of these libraries was derived and discuss the main differences between them.
\par
The \citet{2001ApJ...556..562C} template library consists of 105 templates based on the SEDs of four prototypical starburst galaxies (Arp220 (ULIRG); NGC 6090 (LIRG); M82 (starburst); and M51 (normal star-forming galaxy)). They were derived using the \citet{1998ApJ...509..103S} models with the mid-infrared region replaced with ISOCAM observations between 3 and 18$\mu$m (verifying that the observed values of these four galaxies were reproduced by the templates). These templates were then divided into two portions (4--20$\mu$m and 20--1000$\mu$m) and interpolated between the four to generate a set of libraries of varying shapes and luminosities. The \citet{2001ApJ...549..215D} templates are also included in this set to extend the range of shapes.
\par
The \citet{2002ApJ...576..159D} templates are updated versions of the \citet{2001ApJ...549..215D} templates. This model involves three components, large dust grains in thermal equilibrium, small grains semistochastically heated, and stochastically heated PAHs. They are based on IRAS/ISO observations of 69 normal star-forming galaxies in the wavelength range 3--100$\mu$m. \citet{2002ApJ...576..159D} improved upon these models at longer wavelengths using SCUBA observations of 114 galaxies from the Bright Galaxy Sample (BGS, see \citealt{1989AJ.....98..766S}), 228 galaxies observed with ISOLWS (52--170$\mu$m; Brauher 2002), and 170$\mu$m observations for 115 galaxies from the ISOPHOT Serendipity Survey (\citealt{2000A&A...359..865S}). All together, these 64 templates span the IR luminosity range $10^8-10^{12}L_\odot$.
The total infrared template sample used in our analysis is composed of 168 templates.

\section{Bolometric luminosities and bolometric corrections}
\label{Bolometric luminosities and bolometric corrections}
The nuclear bolometric luminosities and bolometric corrections are estimated, using an approach similar to Pozzi et al. (2007, see also \citealt{2010MNRAS.402.1081V,2010A&A...517A..11P}), whereas the infrared luminosity is used as a proxy of the intrinsic nuclear luminosity. 
The appropriate nuclear template from \citet{2004MNRAS.355..973S} is selected based on the absorbing column density $N_{\rm H}$, when available, or from the best-fit nuclear infrared template.
In order to compute the hard X--ray bolometric correction we used the standard definition
\begin{equation}
 \kbol=\frac{\Lbol}{\Lhard}
\end{equation}
where the $\Lhard$ is the intrinsic X--ray luminosity and the bolometric luminosity is computed as the sum of the total infrared and X--ray luminosity 
\begin{equation}
 \Lbol=\Lir+\Lx.
\end{equation}
After performing the SED-fitting, only the nuclear component of the best-fit is integrated. Hence, the total IR luminosity $\Lir$ is obtained integrating the nuclear template between 1 and 1000$\mu$m. To convert this IR luminosity into the nuclear accretion disk luminosity, we applied the correction factors to account for the torus geometry and the anisotropy (see \citealt{2007A&A...468..603P}).
The first correction is parameterized by the covering factor $f$. The covering factor is related to the geometry of the torus that obscures the accretion disk emission in the optical-UV along the line of sight, and its value is estimated from the ratio of obscured/unobscured quasars found by the X--ray background synthesis models (\citealt{2007A&A...463...79G}). This correction factor is $\sim1.5$. This value correspond to a typical covering factor of $f\sim0.67$, consistent with the results based on clumpy torus models (\citealt{2008ApJ...685..160N}).
\par
The anisotropy factor is defined as the ratio of the luminosity of face-on versus edge-on AGN, where the obscuration is a function of the column density $\NH$. 
Therefore, SEDs in \citet{2004MNRAS.355..973S} have been integrated in the 1--30$\mu$m range, after normalizing these SEDs to the same luminosity in the 30--100$\mu$m range. The derived anisotropy values are 1.2--1.3 for $10^{22}<\NH<10^{24}$ and 3--4 for $\NH>10^{24}$. The same values as in \citet{2010MNRAS.402.1081V} are adopted: 1.3 for $10^{22}<\NH<10^{24}$ and 3.5 for $\NH>10^{24}$.
\par
The total X--ray luminosity $\Lx$ is estimated integrating in the 0.5-100 keV range the X--ray SED. We have interpolated the de-absorbed soft and hard X--ray luminosities. Since we are integrating at the rest-frame frequencies, the X--ray SED is extrapolated at higher and lower energies using the estimated X--ray slope, and introducing an exponential cut-off at 200 keV (\citealt{2007A&A...463...79G}, see also Sect.~3.1 in L10).

\section{Robustness of the method}
\label{Calibrating the method}
\begin{figure*}
  \centering
  {\label{fig:lbolcheck}\includegraphics[width=0.3\textwidth]{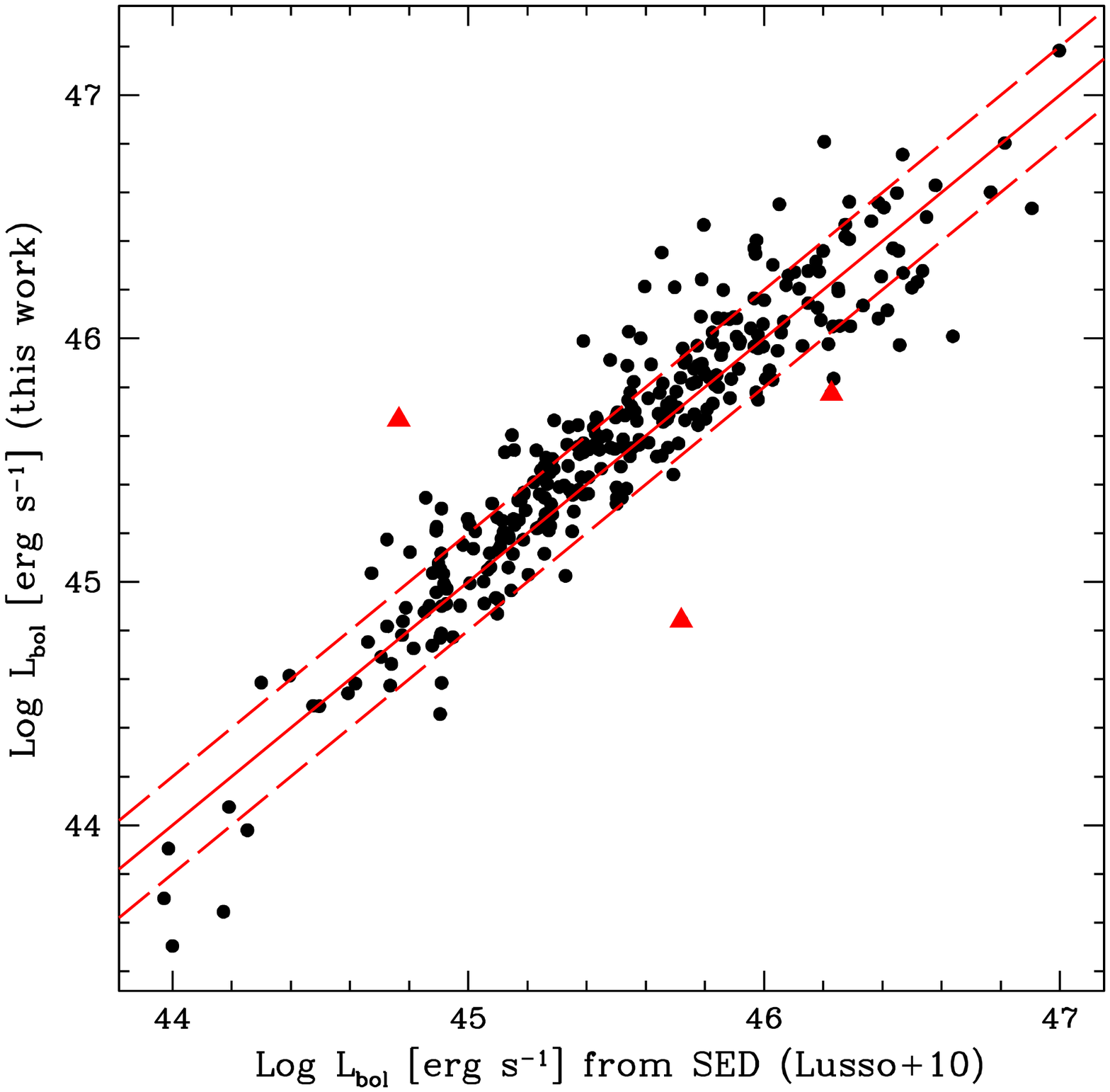}}
  {\label{fig:kbolcheck}\includegraphics[width=0.3\textwidth]{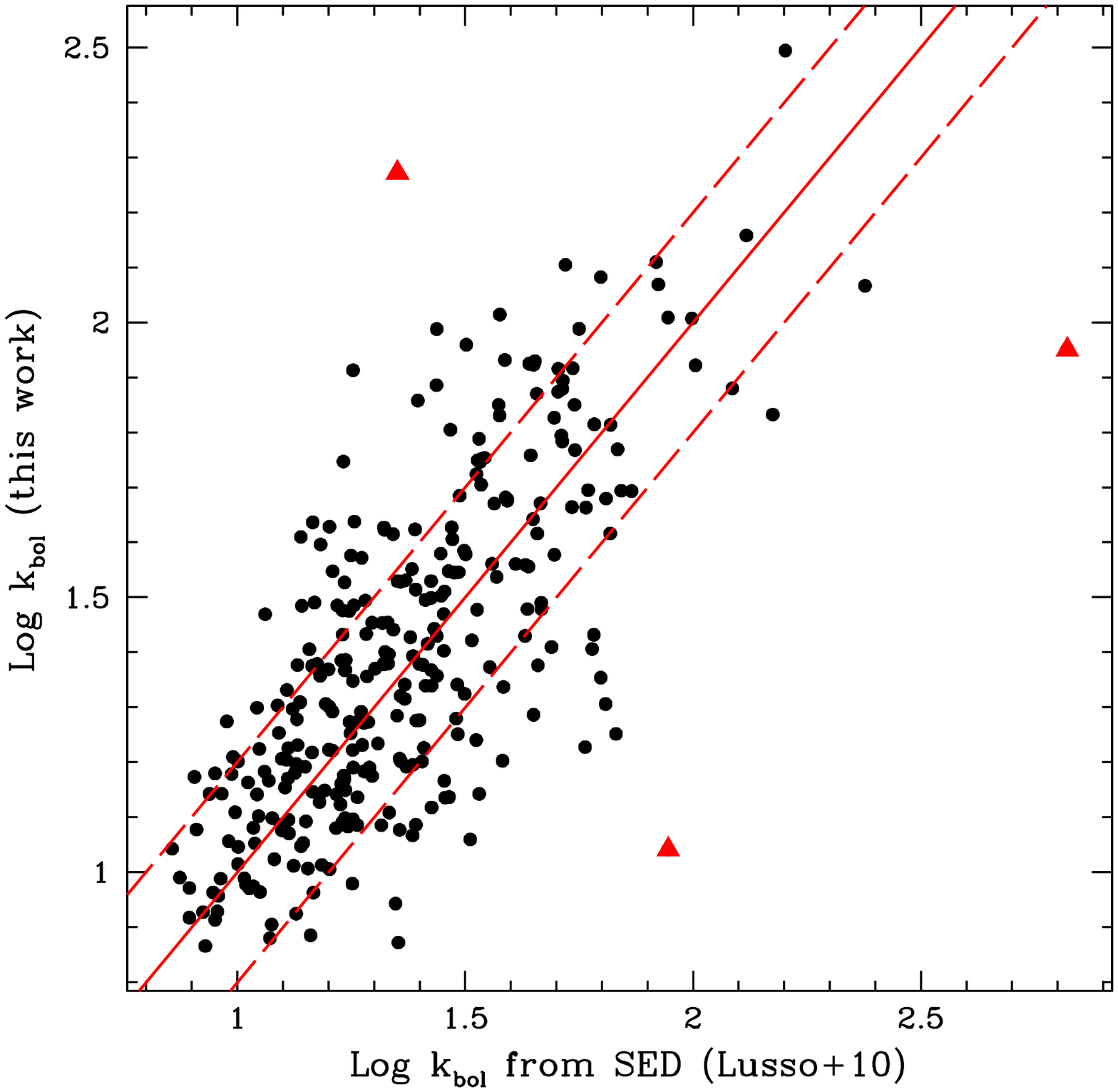}}  \\           
  {\label{fig:histlbolcheck}\includegraphics[width=0.3\textwidth]{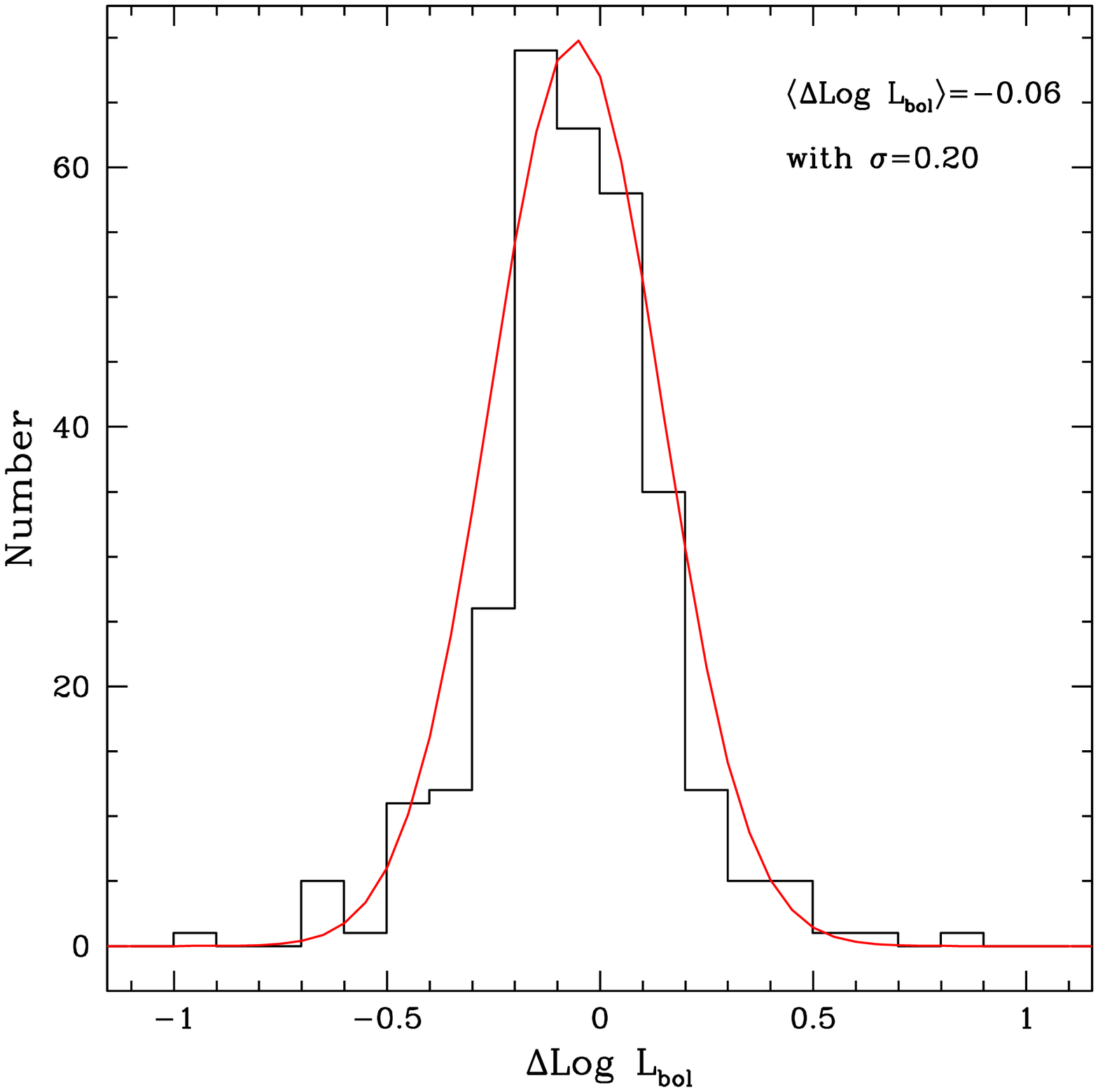}}
  {\label{fig:histkbolcheck}\includegraphics[width=0.3\textwidth]{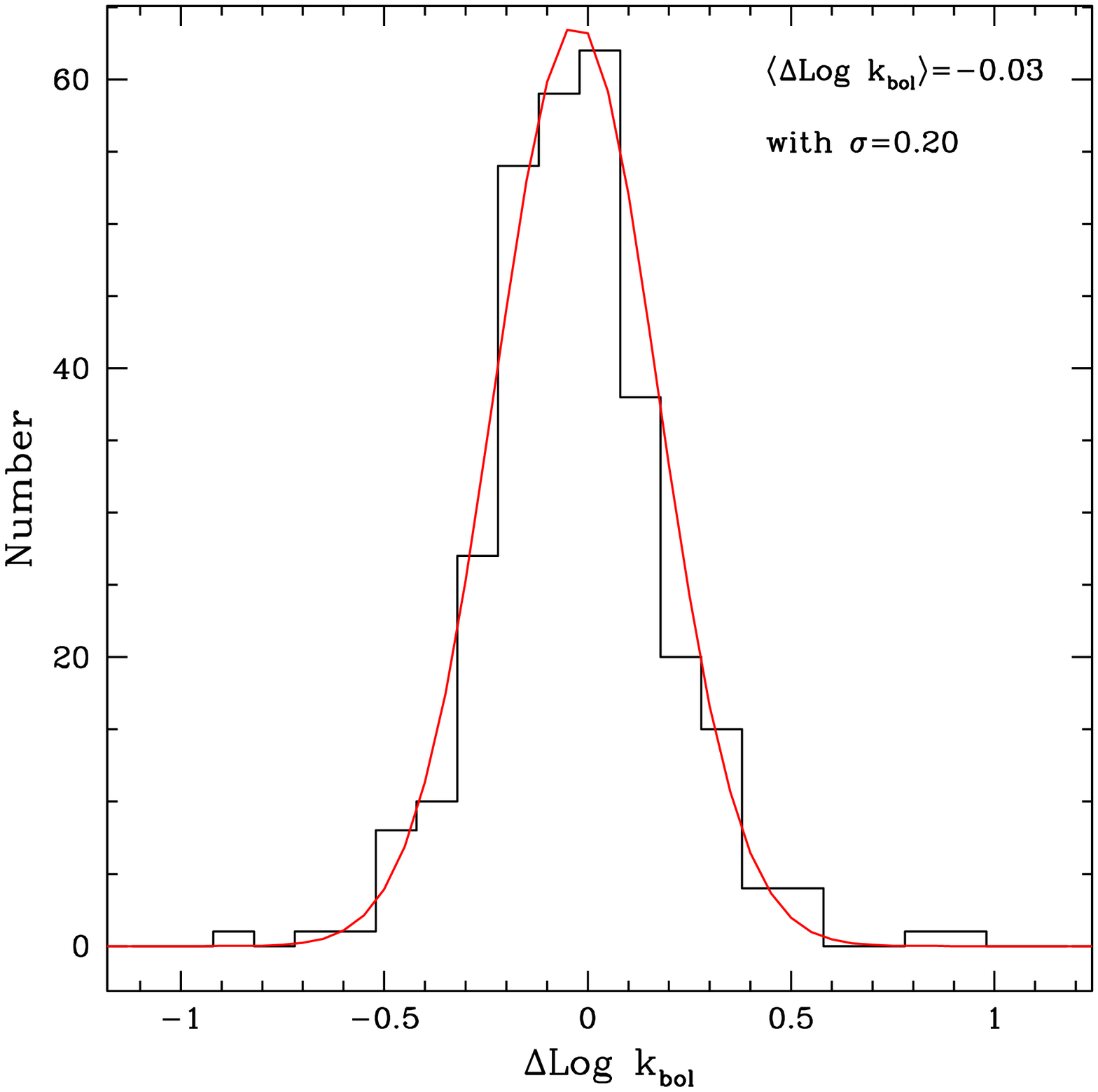}}                
  \caption{Upper panel: Comparison between the values of bolometric luminosity and bolometric correction from data presented in L10 and from this work. The three red triangles mark the outliers discussed in Sect.~\ref{Calibrating the method}. Lower panel: Distribution of the differences between the values of bolometric luminosity and bolometric correction from data presented in L10 and from this work.}
  \label{fig:param}
\end{figure*}
The robustness of the method used to estimate nuclear bolometric luminosities and bolometric corrections from SED-fitting, for the sample of Type-2 AGN, has been tested against the updated soft X--ray selected sample of Type-1 AGN discussed in L10. The Type-1 AGN sample in the L10 work was composed of 361 spectroscopically classified broad-line AGN. The recent work by Brusa et al. (2010) has updated the spectroscopic classification and increased the number of Type-1 AGN with spectroscopic redshift, so that the final sample is composed of 395 Type-1 AGN in the redshift range $0.103\leq z \leq4.255$ with X--ray luminosities $42.20\leq \Log \Lhard\leq 45.23$.
We have computed bolometric and X--ray luminosities, and bolometric corrections using the same approach as in L10 for the Type-1 sample: bolometric luminosites are computed by integrating the rest-frame SEDs from 1$\mu$m up to the UV-bump. In order to compare these estimates with the results from the SED-fitting code, we have applied to the same sample the method described in Sect.~\ref{SED-fitting} and \ref{Bolometric luminosities and bolometric corrections} to estimate bolometric parameters.
To be consistent with the selection criteria of the sample discussed in this paper, we have considered only AGN with X--ray detection in the hard band, removing from the main sample 87 Type-1 AGN with an upper limit at 2--10 keV. Moreover, for 2 sources the best-fit does not consider an AGN component, so we cannot compute the bolometric luminosities for them. 
The final test sample is composed of 306 Type-1 AGN in the redshift range $0.103\leq z \leq3.626$ and X--ray luminosities $42.20\leq \Log \Lhard\leq 45.04$.
\par
In order to select the appropriate nuclear template from \citet{2004MNRAS.355..973S}, we consider the SED for Sy1 AGN (no correction for anisotropy is necessary in this case) and the Sy2 SED with $10^{22}<\NH<10^{23}$ for 20 AGN that have $\NH$ in this range.
\par
We present the comparison between the values of $\Lbol$ and $\kbol$ from L10 and this work in Fig.~\ref{fig:param}.
The outlier in the bottom side of the plot, XID=357 at redshift 2.151 has $\Log \kbol=1.95$ from L10 and $\Log \kbol=1.04$ using the new approach, and presents large error bars in the 24$\mu m$ detection, so that the total bolometric luminosity, computed using the infrared luminosity, is probably underestimated.
The outlier in the right end of the distribution, XID=5114 at redshift 0.212 ($\Log \kbol=2.82$ from L10 and $\Log \kbol=1.95$ using the new approach) has detections at 160, 70 and 24$\mu$m, $\Log \NH = 22.68$ and $\Log \Lhard=42.89$. Probably this source is a star-forming galaxy, so that using the L10 approach we included stellar emission in the estimate of the nuclear bolometric luminosity, thus overestimating the nuclear bolometric luminosity and, therefore, the bolometric correction.
The last notable outlier in the top/left side of the distribution, XID=2152 at redshift 0.627 ($\Log \kbol=1.35$ from L10 and $\Log \kbol=2.27$ using the new approach) presents a significant host-galaxy contribution in the optical-UV and, therefore, the bolometric luminosity is likely to be underestimated in the L10 approach.
\par
Although the two methods are very different, the bolometric luminosity estimates agree remarkably well, with a $1\sigma$ dispersion of 0.20 dex after performing a 3.5$\sigma$ clipping method in order to avoid outliers. Bolometric luminosities from SED-fitting are on average slightly larger than those computed integrating the rest-frame SED from 1$\mu m$ to the X--ray (see the lower left side in Fig.~\ref{fig:param}). 
This effect is also present in the Vasudevan et al. (2010) work. 
A possible explanation is that SED-fitting underestimates the host-galaxy contribution, or that the anisotropy and geometry corrections are too large for some objects.
The agreement between the two methods is overall quite satisfactory and in the following we will discuss our findings for the Type-2 sample.

\section{Results and discussion}
\label{Results and discussion}

\subsection{Bolometric correction and luminosites for Type-2 AGN}

\begin{figure}
 \includegraphics[width=8cm]{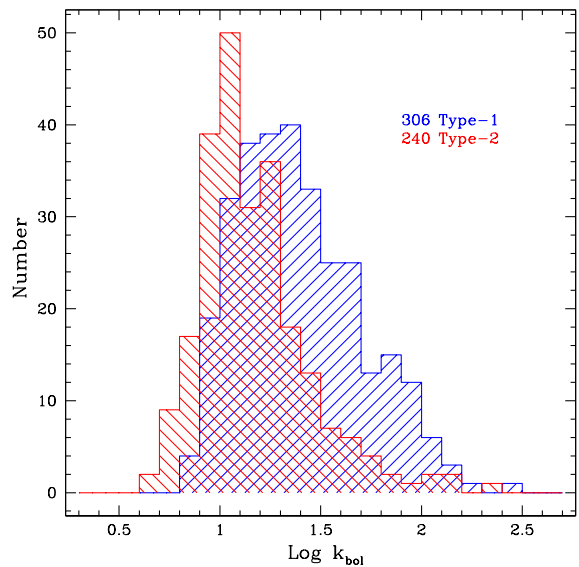}
  \caption{Distribution of the bolometric correction for the 240 Type-2 AGN sample (\textit{red hatched histogram}) and for the 306 Type-1 AGN (\textit{blue hatched histogram}).}
  \label{fig:histkbol12}
\end{figure}

Bolometric luminosities and bolometric corrections have been computed for the Type-2 AGN sample. Intrinsic soft and hard X--ray luminosities are estimated as described in Sect.~\ref{Correction for absorption for the X--ray 2--10 keV luminosity}. For 15 sources we do not have an estimate of the AGN component from the SED-fitting, and we cannot compute the bolometric luminosity for them. 
In Fig.~\ref{fig:histkbol12} the distribution of the bolometric correction for the 240 Type-2 AGN sample and for the 306 Type-1 AGN ($\kbol$ for Type-1 AGN are computed using the SED-fitting code) are presented.
\begin{figure}
 \includegraphics[width=8cm]{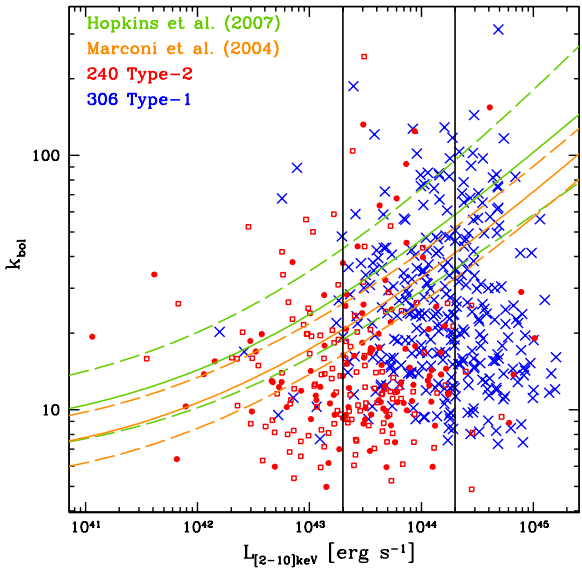}
  \caption{Hard X--ray bolometric correction against the intrinsic 2--10 keV luminosity for 240 Type-2 AGN with AGN best-fit (\textit{red data}). The crosses represent the bolometric correction for 306 Type-1 AGN, computed with the approach described in Sect.~\ref{Bolometric luminosities and bolometric corrections}. The green and blue lines represent the bolometric correction and the $1 \sigma$ dispersion obtained by \citet{hopkins07} and \citet{marconi04}, respectively. The red points and open squares represent the 111 Type-2 AGN with $\NH$ from spectral analyses and the 144 Type-2 AGN with $\NH$ from HR, respectively.}
  \label{fig:l210kbol}
\end{figure}
Figure~\ref{fig:l210kbol} shows bolometric corrections for both the Type-1 and the Type-2 AGN samples as a function of the hard X--ray luminosity. For both samples, bolometric parameters are estimated from the SED-fitting as discussed in Sect.~\ref{SED-fitting}. 
The green and orange curves represent the bolometric corrections and their $1~\sigma$ dispersion as derived by \citet{hopkins07} and \citet{marconi04}, respectively.
Type-2 AGN have, on average, smaller bolometric corrections than Type-1 AGN at comparable hard X--ray luminosity.
For example, at $43.30\leq\Log\Lhard \leq44.30$ (vertical lines in Fig.~\ref{fig:l210kbol}), where both AGN types are well represented, the median bolometric correction for the Type-2 AGN (134 objects) is $\langle \kbol\rangle\sim13\pm1$, to be compared with a median bolometric correction $\langle \kbol\rangle\sim23\pm1$ for the Type-1 AGN (167 objects).
The two averages are statistically different at the $\sim7~\sigma$ level and this is consistent with the results in \citet{2010MNRAS.402.1081V}. The mean $\Lhard$ for the Type-1 and Type-2 AGN within this luminosity range differs by a factor 1.8, and this could in principle explain at least part of the difference in the average bolometric corrections for the two samples of AGN. However, the significance of the difference is still present if we split this luminosity range in two equal $\Log \Lhard$ bins and perform a Kolmogorov--Smirnov test for the Type-1 and Type-2 AGN luminosity distributions in each bin.
\par
\citet{vasudevanfabian09} and \citet{2010A&A...512A..34L} have shown that hard X--ray bolometric corrections are correlated with the Eddington ratios ($\lambda_{\rm Edd}=\Lbol/L_{\rm Edd}$) for Type-1 AGN (see also \citealt{marconi04,kelly08}). The $k_{\rm bol}-\lambda_{\rm Edd}$ relation suggests that there is a connection between the broad-band emission, mostly in the optical-UV, and the Eddington ratio, which is directly linked to the ratio between mass accretion rate and Eddington accretion rate. A high $\lambda_{\rm Edd}$ corresponds to an enhanced optical-UV emission, which means a prominent big-blue bump and therefore a higher $k_{\rm bol}$.
The difference between the average bolometric corrections for Type-1 and Type-2 AGN could be due to lower mass accretion rates in Type-2 AGN, assuming the same black hole mass distribution for the two AGN populations (see \citealt{2011arXiv1103.0276T}). 
The current theoretical framework of AGN/host-galaxy co-evolution predicts that obscured AGN are highly accreting objects and their black hole is rapidly growing. 
However, we note that this is true for $z=1-3$ (see \citealt{2008A&A...490..905H}), while the majority of Type-2 AGN in the present sample are relatively luminous ($42.37\leq\Log \Lbol\leq46.80$), and at moderately low redshift ($0<z<1$). Therefore, the Type-2 AGN sample is likely to represent a later stage in AGN evolution history.

\subsection{Infrared emission: indication of AGN activity}

\begin{figure*}
\centering
 \includegraphics[width=13cm,height=13cm]{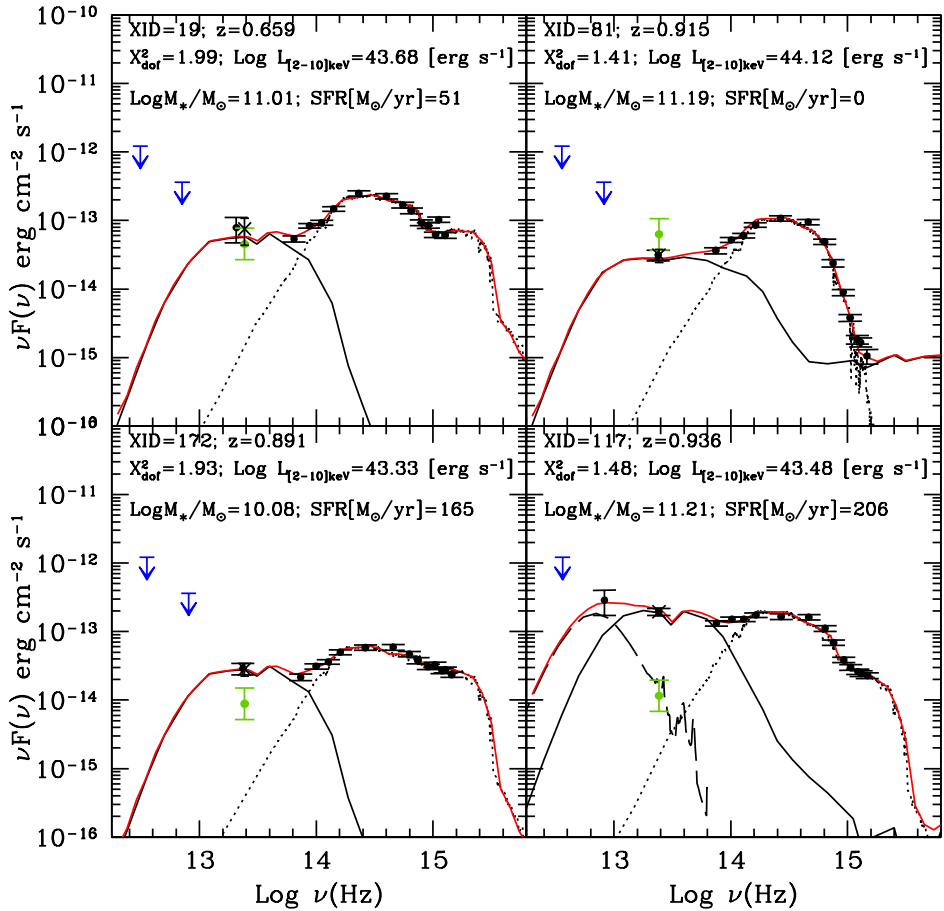}
 \caption{Examples of SED decompositions. Black circles are the observed photometry in the rest-frame (from the far-infrared to the optical-UV). The long-dashed, solid and dotted lines correspond respectively to the starburst, AGN and host-galaxy templates found as the best fit solution. The red line represents the best-fit SED. The stellar mass and the SFR derived from the galaxy template are reported. The green point represents the nuclear mid-infrared luminosity using Eq.~(\ref{gandhieq}), while the cross represents the total observed luminosity at 12.3~$\mu$m computed from the rest-frame SED. XID=19 and 81 are examples of low-$r$ AGN, while XID=172 and 117 represent high-$r$ AGN.}
 \label{sedgandhi}
\end{figure*}
\begin{figure}
 \includegraphics[width=8cm]{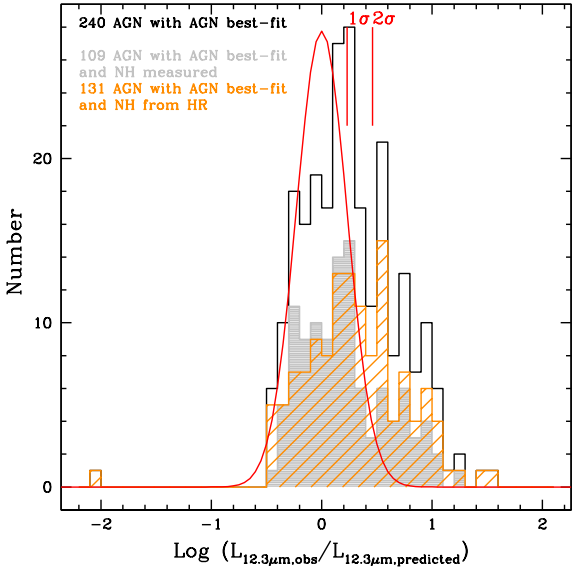}
  \caption{Histogram of the ratio between the total observed luminosity at 12.3~$\mu$m and the mid-infrared luminosity predicted by Eq.~(\ref{gandhieq}). The red curve represents a gaussian with mean equal to zero and standard deviation 0.23. The $1\sigma$ and $2\sigma$ standard deviations of the correlation are also reported.}
  \label{fig:hist_l12.3micron_obspredicted}
\end{figure}

The re-processed infrared emission can be used as a proxy of the average disc emission, since the timescale for transfer of energy from the disk to the outer edge of the torus into infrared emission is of the order of several years in standard AGN picture; whereas optical, UV and X--ray variability in AGN is known to occur on shorter timescales. 
The correlation between the 2--10 keV X--ray emission and IR emission at 12.3$\mu$m for a sample of Seyfert nuclei has been discussed in Gandhi et al. (2009), and it could be used to estimate the intrinsic AGN power. Using X--ray data from the literature and new IR data from the Very Large Telescope's Imager and Spectrometer for mid-Infrared (VISIR), taken specifically for addressing the issue of nuclear emission in local Seyferts, they found a tight correlation between intrinsic, uncontaminated IR luminosity and X--ray luminosity in the 2--10 keV range
\begin{equation}
 \label{gandhieq}
 \Log \frac{L_{12.3~\mu m}}{10^{43}}=(0.19\pm0.05)+(1.11\pm0.07)\Log\frac{L_{[2-10]\rm{keV}}}{10^{43}}.
\end{equation}
The relation is characterized by a small scatter with a standard deviation of 0.23 dex.
The expected nuclear mid-infrared luminosity is computed from Eq.~(\ref{gandhieq}) using the estimate of the intrinsic unabsorbed X--ray luminosity. 
From the observed rest-frame SED (AGN+host-galaxy) the luminosity at 12.3~$\mu$m is computed.
A comparison of the total observed luminosity at 12.3~$\mu$m and that predicted by Eq.~(\ref{gandhieq}) is plotted in Figure~\ref{sedgandhi} for four representative sources. In Figure~\ref{fig:hist_l12.3micron_obspredicted} the distribution of the ratio $r=\Log\left(L_{12.3~\mu m,{\rm obs}}/L_{12.3~\mu m,{\rm predicted}}\right)$ is plotted for the Type-2 AGN sample.
The distribution of the ratio $r$ has a mean which is shifted from zero by $\sim0.2$. 
However, if we consider a gaussian distribution centered at $r=0$ with $\sigma=0.23$, i.e. the same dispersion observed by \citet{2009A&A...502..457G} in their local sample, the majority of the objects are found within $2\sigma$ of the $r$ distribution. 
The tail outside $2\sigma$ and extending to high $r$ includes 73 sources (with $r\gtrsim0.5$) for which the predicted mid-infrared luminosity is significantly lower than observed. 
The hard X--ray luminosities of these 73 AGN are mainly in range $\Log \Lhard\sim42-44$, where the local correlation is well-sampled.
There are two possible explanations for a significant ($\sim30\%$ of the objects) tail toward high-$r$ values: either the Gandhi relation, which was derived for a sample of local Seyfert galaxies, cannot be extended to all the sources in our sample or the SED-fitting procedure may overestimate, in a fraction of these objects, the nuclear contribution. 
In order to study the properties of these outliers, bolometric corrections, morphologies, stellar masses and SFR are discussed in following.
We call ``low-$r$'' AGN all sources within $2\sigma$ of the $r$ distribution, while the ``high-$r$'' AGN sample is populated by the sources deviating more than $2\sigma$ (see Fig.~\ref{sedgandhi} for some examples).
\par
\begin{figure}
 \includegraphics[width=8cm]{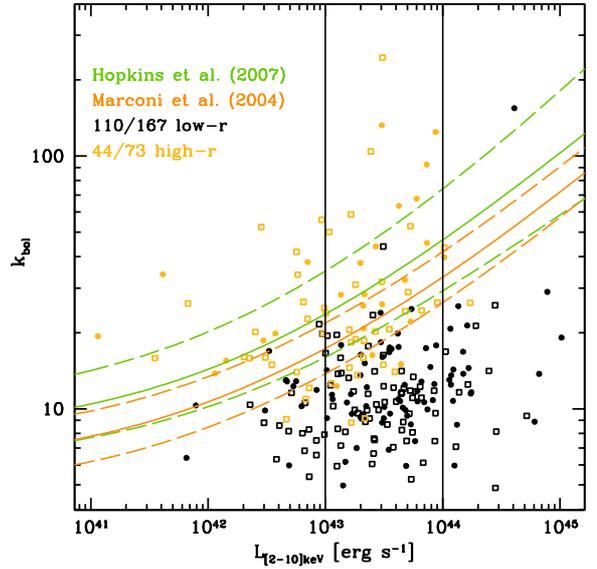}
  \caption{Hard X--ray bolometric correction against 2--10 keV luminosity for 240 Type-2 AGN with AGN best-fit. The 240 Type-2 sample is divided into subsamples: low-$r$ AGN sample (\textit{black data}) and high-$r$ AGN sample (\textit{yellow data}). The green and orange lines represent the bolometric correction and $1\sigma$ dispersion obtained by \citet{hopkins07} and \citet{marconi04}, respectively. In the de-absorbed hard X--ray luminosity range highlighted by the solid lines, we have 167 low-$r$ and 73 high-$r$ (in the infrared) sources.}
  \label{fig:lxkbol_gdad}
\end{figure}

A clear separation in bolometric corrections for these two sub-samples is found. This is shown in Figure~\ref{fig:lxkbol_gdad} in which bolometric corrections are plotted as a function of the 2--10 keV luminosity. At a given hard X--ray luminosity ($43\leq\Log\Lhard \leq44$) the low-$r$ sample has a median bolometric correction of $\langle \kbol\rangle\sim11\pm1$ (110 objects), to be compared with a median bolometric correction for the high-$r$ sample of $\langle \kbol\rangle\sim26\pm3$ (44 objects).
The two median values for $\kbol$ are statistically different at the $\sim5\sigma$ level.  

\par
Furthermore, in the high-$r$ sample 24 Type-2 AGN out of 73 have a detection at 70$\mu$m ($\sim33\%$, significantly higher than those for the low-$r$ sample, $\sim4\%$) and 9 of these 24 have also a detection at 160$\mu$m ($\sim12\%$ considering the total high-$r$ sample, and only $\sim1\%$ for the low-$r$ sample). This denotes that the difference in the average bolometric corrections between the low-$r$ and high-$r$ samples is probably due to the fact that a significant fraction of the infrared emission is attributable to an incorrect modeling of the star-formation process, or the AGN contribution is somehow overestimated by the SED-fitting procedure.
\par
There is no significant difference in the average nuclear absorption between the low-$r$ and the high-$r$ sample, while there is a possibly significant difference in SFR and stellar masses. The median stellar mass in the high-$r$ sample is $\langle \Log M_\ast \rangle\sim 10.93 M_\odot$ with a dispersion of 0.30, while for the low-$r$ sample is $\langle \Log M_\ast \rangle\sim 10.76 M_\odot$ with $\sigma=0.30$. The two averages are statistically different at the $\sim3~\sigma$ level.
The median SFR, as derived from the SED-fit, for the high-$r$ sample is $\langle SFR \rangle\sim 17\pm3 M_\odot$/yrs with a $\sigma=0.30$, while for the low-$r$ sample is $\langle SFR \rangle\sim 3\pm1 M_\odot$/yrs with a $\sigma=0.30$ and the two averages are statistically different at the $4.4~\sigma$ level.
\par
Overall, the SED-fitting for the 73 Type-2 AGN is likely to overestimate the AGN emission in the infrared, which is probably due to the infrared emission from star-forming regions. 
The average bolometric correction for Type-2 AGN, excluding these sources, would be even lower than what we have computed in the previous Section. 
This reinforces the idea of lower bolometric corrections for Type-2 AGN with respect to Type-1 AGN. 
The low bolometric corrections for Type-2 AGN could be also explained if a fraction of the accretion disk bolometric output is not re-emitted in the mid-infrared, but rather dissipated (e.g., by AGN-feedback). This would not be accounted in the bolometric luminosity, and could provide a plausible explanation for low $\kbol$ especially if the low-$r$ sample is considered.
At this stage, this is just a speculation, and more work is needed to verify this possibility.
\begin{table*}
\caption{Properties of the Type-2 AGN sample. \label{tbl_ch4-2}}
\begin{center}
  
\begin{tabular}{lccclllcccc}
\hline \hline
XID & Redshift & $\Log \Lhard$ & $\Log\Lbol$ & $\kbol$ & $\Log M_\ast$ & SFR & $M_{\rm U}$ & $M_{\rm V}$ & $M_{\rm J}$ & Morphological class$^{\mathrm{a}}$ \\
{} & {}  & $\rm{[erg\,s^{-1}]}$ & $\rm{[erg\,s^{-1}]}$ & {} & $\rm{[M_\odot]}$ & $[M_\odot/{\rm yrs}]$ & {} & {} & {} & {} \\
\hline\noalign{\smallskip}  
     67   &   0.367   &   42.79  &    43.80    &  10.24   &    9.68   &    1.86  &   -18.63  &   -19.80   &  -20.72  &     0  \\
     65   &   0.979   &   43.83  &    44.85    &  10.62   &   10.19   &   38.60  &   -20.50  &   -21.63   &  -22.71  &     7  \\
     64   &   0.686   &   43.54  &    44.52    &   9.57   &   10.45   &   32.82  &   -20.17  &   -21.56   &  -23.05  &    10  \\
     63   &   0.355   &   42.98  &    44.28    &  19.88   &   10.79   &    4.94  &   -20.61  &   -22.13   &  -23.26  &     2  \\
     54   &   0.350   &   42.58  &    43.38    &   6.32   &   11.18   &    0.11  &   -20.57  &   -22.62   &  -23.78  &    12  \\
     45   &   0.121   &   41.90  &    42.91    &  10.31   &    9.39   &    0.00  &   -16.76  &   -18.70   &  -19.77  &     1  \\
     43   &   1.162   &   44.22  &    45.28    &  11.49   &   11.30   &    1.47  &   -21.55  &   -23.48   &  -24.61  &     2  \\
     19   &   0.659   &   43.67  &    44.76    &  12.44   &   11.07   &   35.88  &   -21.57  &   -22.88   &  -24.02  &     1  \\
    117   &   0.936   &   43.47  &    45.59    & 132.11   &   11.24   &  199.40  &   -21.81  &   -23.34   &  -24.80  &    23  \\
    116   &   0.874   &   43.49  &    44.50    &  10.17   &   10.56   &    3.77  &   -20.19  &   -21.81   &  -22.90  &     0  \\
    112   &   0.762   &   43.65  &    44.65    &   9.97   &   10.93   &    0.62  &   -20.00  &   -22.13   &  -23.52  &     3  \\
    104   &   0.623   &   44.08  &    45.11    &  10.75   &   10.79   &    0.45  &   -20.25  &   -22.18   &  -23.30  &     1  \\
    101   &   0.927   &   43.70  &    44.68    &   9.56   &   10.92   &    0.61  &   -21.19  &   -22.93   &  -23.78  &     0  \\
    100   &   0.270   &   42.61  &    43.54    &   8.59   &   10.21   &    0.00  &   -19.14  &   -20.98   &  -21.92  &     0  \\
     99   &   0.730   &   43.54  &    44.78    &  17.47   &    9.97   &  116.26  &   -21.61  &   -22.26   &  -23.15  &    11  \\
     85   &   1.001   &   43.46  &    44.84    &  23.99   &   10.13   &   38.52  &   -20.15  &   -21.34   &  -22.56  &     8  \\
     81   &   0.915   &   44.11  &    45.05    &   8.65   &   11.18   &    0.00  &   -20.35  &   -22.59   &  -24.07  &     2  \\
     70   &   0.688   &   44.00  &    45.60    &  39.74   &   10.65   &  542.60  &   -20.87  &   -22.32   &  -24.31  &     3  \\
    152   &   0.895   &   43.75  &    44.82    &  11.72   &    9.87   &   92.88  &   -21.06  &   -21.82   &  -22.85  &     7  \\
    150   &   0.740   &   43.29  &    44.25    &   9.27   &   10.64   &    0.55  &   -19.38  &   -21.28   &  -22.50  &     3  \\
\hline
\end{tabular}
\flushleft \begin{list}{}{Notes---This table is presented entirely in the electronic edition; a portion is shown here for guidance. } 
 \item[$^{\mathrm{a}}$]{The morphological classification of the Type-2 AGN hosts is coded from 0 to 23: 0 = elliptical, 1 = S0; 2 = bulge-dominated; 3 = intermediate-bulge; 4 = disk-dominated; 5 = irregular; 6 = compact/irregular; 7 = compact; 8 = unresolved/compact; 9 = blended; 10 = bulge-dominated/close-companion; 11 = intermediate-bulge/close-companion; 12 = S0/close-companion; 23 = possible mergers.}
 \end{list}
\end{center}

\end{table*}

\subsection{Host-galaxy properties: $M_\ast$, SFR, colors and morphologies}
\begin{figure}
 \includegraphics[width=8cm]{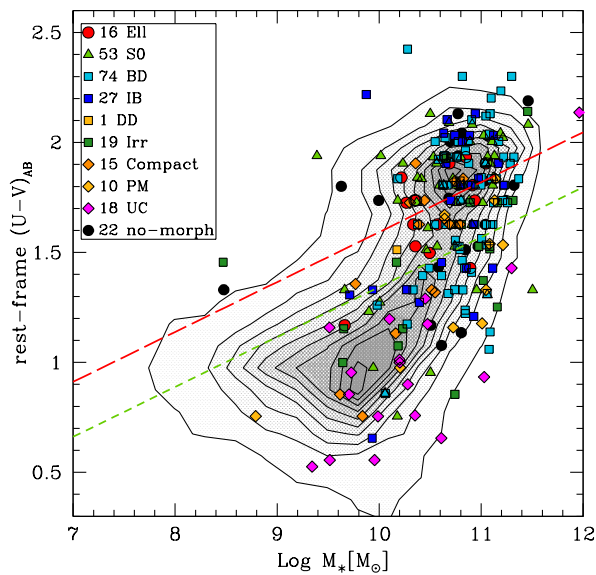}
 \caption{The morphology distribution (using the ZEST+ code) of the 233 AGN host-galaxies on the $(U-V)$ colour-mass diagram. We also plotted the 22 sources without morphological information. The $(U-V)$ color and stellar masses are computed using the SED-fitting code. We overplot the contours of about 8700 galaxies in $z$COSMOS (colours and stellar masses from \textit{Hyperz}). The morphology classification is labeled as follow: elliptical (Ell), S0, bulge-dominated galaxy (BD), intermediate-bulge galaxy (IB), disk-dominated galaxy (DD), irregular (Irr), Compact, possible mergers (PM) and unresolved compact (UC). The red dashed line represents the red sequence cut defined by \citet{2006A&A...453..869B}, while the green short dashed line defines an approximate green valley region, both lines are calculated at redshift $\sim 0.76$, which is the average redshift of the main Type-2 sample.}
 \label{contour_morp_uv_zest+}
\end{figure}
Galaxies show a colour bi-modality both in the local Universe and at higher redshift (up to $z\sim2$; e.g., \citealt{2001AJ....122.1861S,2004ApJ...608..752B}). This bi-modality (red-sequence and blue-cloud galaxies) has been interpreted as an evidence for a dicothomy in their star formation and merging histories (e.g., \citealt{2005ApJ...632...49M}, but see also \citealt{2010ApJ...721L..38C} for an alternative explanation). Color-magnitude and color-mass diagrams  (e.g., rest-frame $(U-V)$ versus stellar mass) have been used as tools in galaxy evolution studies, and since many models invoke AGN feedback as an important player in such evolution, it is interesting to locate the hosts of Type-2 AGN in those diagrams. Using the galaxy component obtained from the best fit of the Type-2 AGN, it is possible to derive rest-frame colors for the host that, linked to the stellar mass and the morphology, can provide hints on AGN feedback. Several studies found that the hosts of obscured AGN tend to be redder than the overall galaxy population in the rest-frame $(U-V)$ color (e.g., \citealt{2007ApJ...660L..11N}). There are at least two possible and significantly different interpretations for this observational result: the observed red colors are mainly due to dust extinction, so that a significant fraction of obscured AGN would live in massive, dusty star-forming galaxies with red optical colors (e.g., \citealt{2009A&A...507.1277B}); or red sources are linked with passive systems (e.g., \citealt{2007A&A...475..115R,2009ApJ...692L..19S,2010ApJ...721L..38C}).
Therefore, accurate stellar mass and SFR estimates, together with detailed galaxy morphologies, are of particular importance to discriminate between the two alternative possibilities.
\par
The very high resolution and sensitivity  of ACS-HST imaging in the COSMOS survey provides resolved morphologies for several hundreds of thousands galaxies with $i_{\rm acs}\leq 24$ (see Scarlata et al 2007 for details).
Galaxy morphologies were obtained with an upgraded version of the Zurich Estimator of Structural Types (ZEST; \citealt{2007ApJS..172..406S}), known as ZEST+ (Carollo et al. 2011, in prep). Relative to its predecessor, ZEST+ includes additional measurements of non-parametric morphological indices for characterising both structures and substructures. For consistency with the earlier versions, ZEST+ uses a Principal Component Analysis (PCA) scheme in the 6-dimensional space of concentration, asymmetry, clumpiness, M$_{20}$ (second-order moment of the brightest 20\% of galaxy pixels), Gini coefficient, and ellipticity. ZEST+ classifies galaxies in seven morphological types located in specific regions of the 6-dimensional space: elliptical, S0, bulge-dominated disk, intermediate-bulge disk, disk-dominated, irregular, compact. The different types were then visually inspected. For 19 objects ZEST+ is unable to give any information on morphology because these sources lie off the edge of the ACS tiles and 4 sources are blended.
As a result of the ZEST+ procedure and visual inspection of the other 233 galaxies in our sample, we find that 16 are ellipticals (Ell), 53 are S0s, 74 are bulge-dominated (BD) disks, 27 are intermediate-bulge (IB) disks, just 1 is disk-dominated (DD), 19 are irregular galaxies (Irr), 15 are compact galaxies (i.e. the structural parameters computed for these galaxies from the HST-ACS images are highly affected by the instrumental PSF) and 18 are unresolved compact galaxies (UC, i.e. essentially point-like sources). Ten galaxies show distortions and potential signatures of ongoing or recent mergers (PM).
\rev{At the typical magnitudes of the objects in our sample, the ZEST+ classification is highly reliable for galaxies with redshift $\lesssim1$. At higher redshifts, morphological k-correction and, to a lesser extent, resolution effects can adversely affect measurements of ZEST+ parameters (note that only 4\% of the main Type-2 AGN sample have $z>1.5$). However, broad morphological bins (e.g., early/late type galaxies) should be relatively robust. For high-$z$ galaxies, resolution might also have an impact on the classification for mergers, and ACS images could not be deep enough to distinguish merger features (see also \citealt{2011arXiv1105.5395M}). Moreover, inclination might also affect the morphological classification (e.g., \citealt{2010ApJS..186..427N}). However, a detailed study of systematics and biases in the morphological classification is beyond the purposes of the present paper.}
In Table \ref{tbl_ch4-2} we list the main properties of the sample.
\par
The rest-frame $(U-V)$ color encompasses the 4000$\text{\AA}$ break, and it is particularly sentitive to age and metallicity variations of stellar population in galaxies (e.g., \citealt{1978ApJ...225..742S,2004ApJ...608..752B,2006A&A...453..869B,2010ApJ...721L..38C}).
In Fig.~\ref{contour_morp_uv_zest+} the distribution of the rest-frame $(U-V)$ colors, which are computed directly from the best-fit galaxy template, and stellar masses (from the SED-fitting code) are reported for the entire Type-2 AGN sample. In the same figure, the background contours for a sample of $\sim8700$ galaxies in zCOSMOS ($i_{\rm acs}<22.5$, 240 Type-2 are detected in the $i_{\rm acs}$ band, 183/240 Type-2 AGN (76\%) have $i_{\rm acs}<22.5$) are also plotted, where colours and stellar masses are computed using the Hyperz code (\citealt{2000A&A...363..476B}).
\par
AGN are known to reside in massive galaxies (e.g., \citealt{2009ApJ...696..396S,2009A&A...507.1277B}) and this is fully confirmed by the present analysis. 
The morphologies of the host-galaxies and the stellar masses indicate that there is a preference for these Type-2 AGN to be hosted in bulge-dominated and S0 galaxies ($\sim50\%$) with stellar masses greater than $10^{10}M_\odot$. 
This result is consistent with the previous studies on Type-2 AGN by \citet[see also \citealt{2003MNRAS.346.1055K,2008ApJ...681..931B,2011ApJ...727L..31S}]{2008ApJ...675.1025S}.
\par
It should be noted that no correction for the internal extinction has been applied to the $(U-V)$ colors of both background galaxies in zCOSMOS and Type-2 AGN hosts. This correction could be important as shown in \citet{2008ApJ...686...72C} (see also \citealt{2010ApJ...721L..38C}). In that work star formation and galactic stellar mass assembly are analyzed using a very large and highly spectroscopically complete sample selected in the rest-frame NIR bolometric flux in the GOODS-N. They found that applying extinction corrections is critical when analyzing galaxy colors; nearly all of the galaxies in the green valley are 24$\mu m$ sources, but after correcting for extinction, the bulk of the 24$\mu m$ sources lie in the blue cloud.
This correction introduces an average shift in color of $\sim0.2$~mag for the most extincted/star-forming galaxies. 
However, to be consistent with the colors of the background galaxies no correction for intrinsic extinction is considered.
\par
\begin{figure}
 \includegraphics[width=8cm]{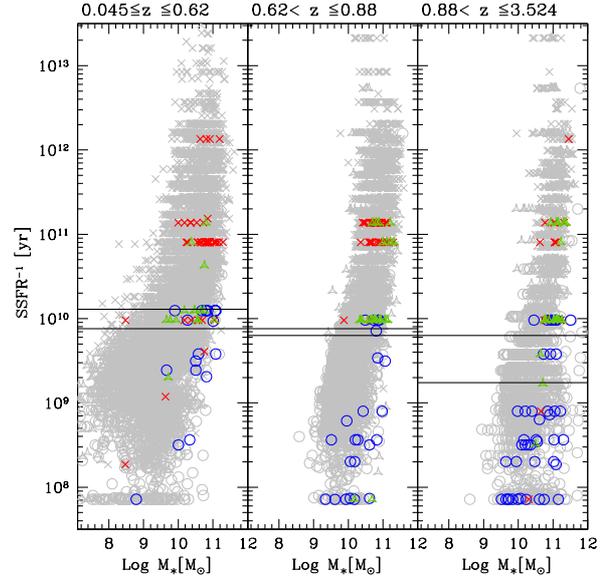}
 \caption{Inverse of the SSFR rate as a function of the stellar mass of the AGN host-galaxies in three different redshift bins for the zCOSMOS galaxies and for the Type-2 AGN sample in the red-sequence (red crosses), in the green-valley (green triangles) and in the blue-cloud (blue open circles). The horizontal lines mark the age of the Universe at the two redshift boundaries of the chosen intervals.}
 \label{ssfrmstar}
\end{figure}

AGN host-galaxies belong to the red-sequence if their $(U-V)$ color is above the threshold (\citealt{2006A&A...453..869B}):
\begin{equation}
 (U-V)_{\rm AB, rest-frame}>0.277 \Log M_\ast -0.352 z -0.39
\end{equation}
Sources in the green-valley are approximately defined shifting this relation by 0.25 downward towards bluer colors. With these definitions, $\sim42\%$ (108/255) and $\sim25\%$ (63/255) of the total sample are included in the red-cloud and the green-valley, respectively.
For all sources the Specific Star-Formation Rate (SSFR) is estimated, defined as the ratio of the SFR per unit of galaxy stellar mass (SSFR=SFR/$M_\ast$). The inverse of the SSFR, $SSFR^{-1}$, is called ``growth time''
\begin{equation}
\label{ssfr}
 SSFR^{-1}=M_\ast/\dot M_\ast,
\end{equation}
and corresponds to the time required for the galaxy to double its stellar mass, assuming its SFR remained constant. Actively star-forming galaxies are defined as sources with growth time smaller than the age of the Universe at their redshift ($SSFR^{-1} < t_{\rm Hubble}$), while sources with $SSFR^{-1}$ larger than the age of the Universe can be considered passive galaxies (see also \citealt{2009A&A...501...15F,2009A&A...507.1277B}).
Figure~\ref{ssfrmstar} shows $SSFR^{-1}$ as a function of the stellar mass in three different redshift bins for the AGN host-galaxies in the red-sequence, in the green-valley and in the blue-coud and for the zCOSMOS galaxies in same redshift ranges.
The horizontal lines mark the age of the Universe at the two redshift boundaries of the chosen intervals.
At face value, almost all the sources in the red-sequence have $SSFR^{-1}$ larger than the age of the Universe at their redshift, which is consistent with passive galaxies. However, the value of $SSFR^{-1}$ has to be considered only as an approximate indication of the star-formation activity; in fact, there is some possible evidence of some residual star-formation, in red-cloud AGN host-galaxies, as witnessed by their morphologies.
In the red-sequence 8 and 28 sources are classified as ellipticals and S0s, respectively; all together they represent 34\% of the host-galaxy population in the red-sequence. About 42\% is represented by disk galaxies (both bulge-dominated and intemediate-bulge), which are probably still forming stars but not at high rates.
In fact, 15 over 108 sources ($\sim14\%$) have a detection at 70$\mu$m and 5 have also a detection at 160$\mu$m ($\sim6\%$).
\par
\begin{figure}
 \includegraphics[width=8cm]{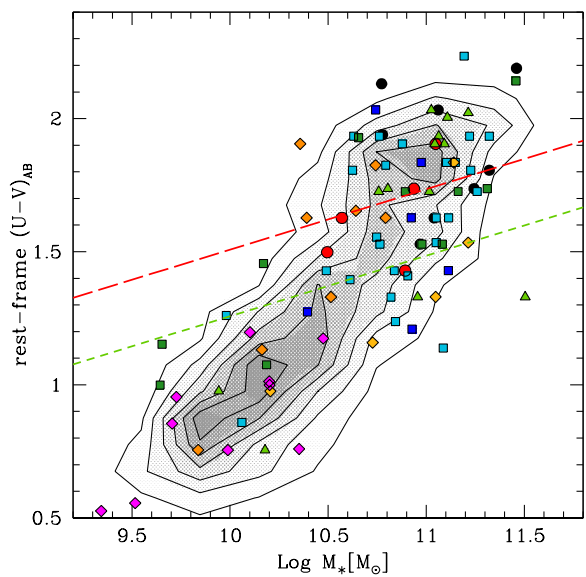}
 \caption{Distribution of the stellar masses as a function of the rest-frame $(U-V)$ colors in the redshift range $0.8\leq z\leq 1.2$. The red dashed line represents the red sequence cut defined by \citet{2006A&A...453..869B}, while the green short dashed line defines an approximate green valley region, both lines are calculated at redshift $\sim 1$. The points are color coded as in Fig.~\ref{contour_morp_uv_zest+}.
}
 \label{mstaruvredshiftsel}
\end{figure}
\begin{figure}
 \includegraphics[width=8cm]{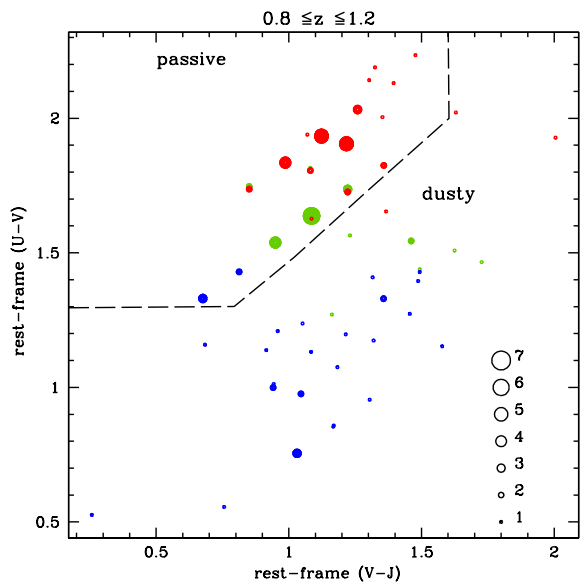}
 \caption{Distribution of Type-2 AGN hosts in the rest-frame $(U-V)$ against the rest-frame $(V-J)$ color. Color coded as in Fig.~\ref{ssfrmstar}. Sources with the same best-fit galaxy template and the same extinction lie in the same position in the color-color diagram. Point size is keyed to the number of objects. 
}
 \label{uvvj}
\end{figure}

For these objects, the SFR inferred from the far-infrared detections is significantly higher than the SFR derived from the SED-fitting procedure. Indeed, an SED-fitting over the UV, optical and near-infrared bands is not always able to discriminate between the red continua of passive galaxies and those of dusty star-forming galaxies.
Therefore, we decided to include another indicator in the present analysis broadly following the procedure described in \citet[i.e., the $(U-V)-(V-J)$ color diagram]{2010ApJ...721L..38C}. 
Near-infrared emission can distinguish between red-passive or dust-obscured galaxies: given a similar $0.5\mu$m flux, a star-forming galaxy has more emission near $\sim1\mu$m than a passive galaxy.
A sub-sample of galaxies is selected in the same redshift range explored by \citet{2010ApJ...721L..38C}, we find 92 AGN host-galaxies with $0.8\leq z\leq 1.2$. Fig.~\ref{mstaruvredshiftsel} shows both inactive galaxies and AGN host-galaxies in the same redshift range and the thresholds considered to divide galaxies in the red-sequence and in the green-valley (we consider an average redshift of 1 to define the threshold for the red-sequence and the green-valley). Thirty-five out of 92 AGN hosts are found to lie in the red-sequence ($\sim38\%$) and 23 in the green-valley ($\sim25\%$); while for inactive galaxies about 32\% and 21\% lie in the red-sequence and green-valley, respectively.
In Figure~\ref{uvvj} the $(U-V)-(V-J)$ color diagram for the 92 Type-2 AGN hosts is presented.  
From a preliminary analysis of the rest frame $(U-V)$ against the rest-frame $(V-J)$ color (see Fig.~\ref{uvvj}, but see also Fig.~2 in \citealt{2010ApJ...721L..38C}), only $\sim9\%$ of the AGN host-galaxies in the red-sequence and $\sim30\%$ of AGN host-galaxies in the green-valley are moved in the region populated by dusty star-forming galaxies in the color-color diagram. To be compared with $20\%$ AGN host-galaxies in the red-sequence and $75\%$ AGN host-galaxies in the green-valley found by Cardamone and collaborators.
The fractions of dust-obscured galaxies among the red-cloud and green-valley AGN in our sample, at $0.8\leq z\leq1.2$, are lower than those in the \citet{2010ApJ...721L..38C} sample. 
However, the global fractions of AGN hosts, tentatively associated to passive galaxies, are very similar ($\sim50\%$) in the two samples.
The fractions for both AGN host-galaxies and inactive galaxies are reported in Table~\ref{tbl-2}.


\begin{table*}[ct]
\caption{AGN hosts and galaxies properties. \label{tbl-2}} 
\centering
\begin{tabular}{l c| cl|cl|c}
Sample & N & Red-sequence & & Green-valley & & Blue-cloud 
\\ [0.5ex]
\hline\hline\noalign{\smallskip}
\multicolumn{7}{c}{$0.045\leq z \leq 3.452$} \\
\hline\noalign{\smallskip}
&& &96 (89\%) P && 52 (82\%) P  \\[-1ex]
\raisebox{1.5ex}{Type-2 AGN} & \raisebox{1.5ex}{255}& \raisebox{1.5ex}{108 (42\%)} & 12 (11\%) D & \raisebox{1.5ex}{63 (25\%)}
& 11 (18\%) D & \raisebox{1.5ex}{84 (33\%)}  \\[1ex]
&& & 2306 (91\%) P && 1596 (83\%) P \\[-1ex]
\raisebox{1.5ex}{Galaxies} & \raisebox{1.5ex}{8742}& \raisebox{1.5ex}{2535 (29\%)} & 229 (9\%) D & \raisebox{1.5ex}{1923 (22\%)}
& 327 (17\%) D & \raisebox{1.5ex}{4284 (49\%)}  \\[1ex]
\hline\hline\noalign{\smallskip}
\multicolumn{7}{c}{$0.8\leq z \leq 1.2$} \\
\hline\noalign{\smallskip}
&& &32 (91\%) P && 16 (70\%) P  \\[-1ex]
\raisebox{1.5ex}{Type-2 AGN} & \raisebox{1.5ex}{92}& \raisebox{1.5ex}{35 (38\%)} & 3 (9\%) D & \raisebox{1.5ex}{23 (25\%)}
& 6 (30\%) D & \raisebox{1.5ex}{34 (37\%)}  \\[1ex]
&& &569 (97\%) P && 269 (70\%) P  \\[-1ex]
\raisebox{1.5ex}{Galaxies} & \raisebox{1.5ex}{1836}& \raisebox{1.5ex}{587 (32\%)} & 18 (3\%) D & \raisebox{1.5ex}{385 (21\%)}
& 116 (30\%) D & \raisebox{1.5ex}{864 (47\%)}  \\[1ex]
\hline\hline
\end{tabular}

\flushleft\begin{list}{}{Note -- P=Passive, D=Dusty.}
 \item
\end{list}
\end{table*}

\section{Summary and Conclusions}
\label{Summary and Conclusions}
A detailed analysis of the SEDs of 255 spectroscopically identified hard X--ray selected Type-2 AGN from the XMM-COSMOS survey is presented. 
In obscured AGN, the optical-UV nuclear luminosity is intercepted along the line of sight by the dusty torus and reprocessed in the infrared, so what we see in the optical-UV is mostly the light from the host-galaxy. On the one hand, this allows us to study the galaxy properties, on the other hand it makes difficult to estimate the nuclear bolometric  power. An SED-fitting code has been developed with the main purpose of disentagling the various contributions (starburst, AGN, host-galaxy emission) in the observed SEDs using a standard $\chi^2$ minimization procedure (the starburst component is only used in the case of detection at 70$\mu$m). The code is based on a large set of starburst templates from \citet{2001ApJ...556..562C} and \citet{2001ApJ...549..215D}, and galaxy templates from the \citet{2003MNRAS.344.1000B} code for spectral synthesis models, while AGN templates are taken from \citet{2004MNRAS.355..973S}. These templates represent a wide range of SED shapes and luminosities and are widely used in the literature.
The total (nuclear) AGN bolometric luminosities are then estimated by adding the X--ray luminosities integrated over the 0.5-100 keV energy range to the infrared luminosity between 1 and 1000$\mu$m.
The total X--ray luminosity is computed integrating the X--ray SED using the de-absorbed soft and hard X--ray luminosities. The SED is extrapolated to higher energies using the observed X--ray slope, and introducing an exponential cut-off at 200 keV. 
The total infrared luminosity is evaluated integrating the infrared AGN best-fit and then converted into the nuclear accretion disk luminosity applying the appropriate correction factors to account for the geometry and the anisotropy of the torus emission.
The reprocessed IR emission is considered to be a good proxy of the intrinsic disk emission and this is supported by previous investigations (\citealt{2007A&A...468..603P}; Gandhi et al. 2009; \citealt{2010MNRAS.402.1081V}). 
In the distribution of the ratio $r=\Log\left(L_{12.3~\mu m,{\rm obs}}/L_{12.3~\mu m,{\rm predicted}}\right.$; see Eq.~\ref{gandhieq}) the majority of the objects are within $2\sigma$ of the $r$ distribution. 
The tail outside $2\sigma$ and extending to high $r$ includes 73 sources (with $r\gtrsim0.5$) for which the predicted mid-infrared luminosity is significantly lower than observed. 
We call ``low-$r$'' AGN all sources within $2\sigma$ of the $r$ distribution, while the ``high-$r$'' AGN sample is represented by the sources deviating more than $2~\sigma$.
\par
Our main results are the following:
\begin{enumerate}
 \item The average observed SED is characterized by a flat X--ray slope, $\langle\Gamma=1.12\rangle$, as expected for obscured AGN (not corrected for absorption), while in the optical-UV the observed light appears to be consistent with the host-galaxy emission. 
The average SED in the mid-infrared is more likely a combination of dust emission from star-forming region and AGN emission reprocessed by the dust.
 \item The full sample is split into four bins of different X--ray and infrared luminosities and redshift. The shapes of the average SEDs in the optical bands are approximately the same in all luminosity and redshift bins. There is a stronger host-galaxy contribution at lower luminosity/redshift bins, where the average SEDs have a typical galaxy shape. Moreover, there is a trend between X--ray and mid-infrared luminosity: the contribution of the AGN in the infrared (around $8-15\mu$m) is higher at higher X--ray luminosities.

 \item Type-2 AGN appear to have smaller bolometric corrections than Type-1 AGN. At the same hard X--ray luminosity, $43.30\leq\Log\Lhard \leq44.30$, where both samples are well represented, we find that the median bolometric correction for Type-2 AGN (134 objects) is $\langle \kbol\rangle\sim13\pm1$, to be compared with a median bolometric correction $\langle \kbol\rangle\sim23\pm1$ for Type-1 AGN (167 objects). The two averages are statistically different at the $\sim7~\sigma$ level.

 \item A clear separation in bolometric corrections for the low-$r$ and the high-$r$ samples is found. The relation provided by Gandhi and collaborators is valid for the majority of objects, while for 30\% of the sample SED-fitting procedure may underestimate the non-nuclear contribution. At a given hard X--ray luminosity ($43\leq\Log\Lhard \leq44$) the low-$r$ sample has a median bolometric correction of $\langle \kbol\rangle\sim11\pm1$ (110 objects), to be compared with a median bolometric correction for the high-$r$ sample of $\langle \kbol\rangle\sim26\pm3$ (44 objects). The two median values for $\kbol$ are statistically different at the $\sim5\sigma$ level.

 \item Host-galaxies morphologies and the stellar masses indicate that Type-2 AGN are preferentially hosted in galaxies which have a bulge, irrespective of the strength of the bulge or if the galaxy is on the red sequence or blue cloud, and with stellar masses greater than $10^{10}M_\odot$.

 \item Almost all the sources in the red-sequence have $SSFR^{-1}$ larger than the age of the Universe at their redshift, which is consistent with passive galaxies. Following the same approach as in Cardamone and collaborators (i.e., combining the rest-frame $(U-V)$ vs $\Log M_\ast$ and the rest-frame $(U-V)$ vs $(V-J)$ color diagrams), we find that, consistent with their results, $\sim50\%$ of AGN hosts lie in the passive region of this diagram. In contrast from Cardamone et al. (2010), only $\sim30\%$ of AGN host-galaxies in the green-valley in our sample are consistent with dust-obscured sources in $0.8\leq z\leq 1.2$.
\end{enumerate}
It is clear that the mid and far-infrared parts of the SED are under-sampled with respect to the optical part. The ongoing Herschel survey over various fields at different depths ($100\mu m$ and $160\mu m$ in the COSMOS field) and the upcoming ALMA surveys will allow us to gain an optimal multiwavelength coverage also in the far-infrared.

\begin{acknowledgements}
We gratefully thank B. Simmons for her useful comments and suggestions.
In Italy, the XMM-COSMOS project is supported by ASI-INAF grants I/009/10/0, I/088/06 and ASI/COFIS/WP3110
I/026/07/0. Elisabeta Lusso gratefully acknowledges financial support from the Marco Polo program, University of Bologna. In Germany the XMM-\textit{Newton} project is supported by the Bundesministerium f\"{u}r Wirtshaft und Techologie/Deutsches Zentrum f\"{u}r Luft und Raumfahrt and the Max-Planck society. Support for the work of E.T. was provided by NASA through Chandra Postdoctoral Fellowship Award grant number PF8-90055, issued by the Chandra X-ray Observatory Center, which is operated by the Smithsonian Astrophysical Observatory for and on behalf of NASA under contract NAS8-03060. The entire COSMOS collaboration is gratefully acknowledged.
\end{acknowledgements}

\bibliographystyle{aa}
\bibliography{bibl}

\end{document}